\documentclass[a4paper,fleqn,usenatbib,useAMS]{mnras}
\usepackage{caption}
\usepackage{subcaption}
\usepackage{graphicx}	\usepackage{amsmath}	\usepackage{amssymb}	\usepackage{multicol}        \usepackage{bm}		\usepackage{pdflscape}	
  
\usepackage[T1]{fontenc}
\usepackage{ae,aecompl}

\title[Radial gradients in Coma BCGs]{Radial gradients in initial mass function sensitive absorption features in the Coma brightest cluster galaxies}
\author[Zieleniewski et al.]{Simon Zieleniewski$^{1}$\thanks{E-mail:
simon.zieleniewski@physics.ox.ac.uk} Ryan C. W. Houghton$^{1}$ Niranjan Thatte$^{1}$ Roger L. Davies$^{1}$\newauthor
Sam P. Vaughan$^{1}$\\
$^{1}$Astrophysics, Denys Wilkinson Building, Keble Road, Oxford, OX1 4RH}
\begin{document}

\date{Submitted 15-10-16}

\pagerange{\pageref{firstpage}--\pageref{lastpage}} \pubyear{2014}

\maketitle

\label{firstpage}

\begin{abstract}
Using the Oxford Short Wavelength Integral Field specTrograph (SWIFT), we trace radial variations of initial mass function (IMF) sensitive absorption features of three galaxies in the Coma cluster. We obtain resolved spectroscopy of the central $5\,\rmn{kpc}$ for the two central brightest-cluster galaxies (BCGs) NGC4889, NGC4874, and the BCG in the south-west group NGC4839, as well as unresolved data for NGC4873 as a low-$\sigma_*$ control. We present radial measurements of the IMF-sensitive features sodium NaI$_{\rm{SDSS}}$, calcium triplet CaT and iron-hydride FeH0.99, along with the magnesium MgI0.88 and titanium oxide TiO0.89 features. We employ two separate methods for both telluric correction and sky-subtraction around the faint FeH feature to verify our analysis. Within NGC4889 we find strong gradients of NaI$_{\rm{SDSS}}$ and CaT but a {\it flat} FeH profile, which from comparing to stellar population synthesis models, suggests an old, $\alpha$-enhanced population with a Chabrier, or even bottom-light IMF. The age and abundance is in line with previous studies but the normal IMF is in contrast to recent results suggesting an increased IMF slope with increased velocity dispersion. We measure flat NaI$_{\rm{SDSS}}$ and FeH profiles within NGC4874 and determine an old, possibly slightly $\alpha$-enhanced and Chabrier IMF population. We find an $\alpha$-enhanced, Chabrier IMF population in NGC4873. Within NGC4839 we measure both strong NaI$_{\rm{SDSS}}$ and strong FeH, although with a large systematic uncertainty, suggesting a possible heavier IMF. The IMFs we infer for these galaxies are supported by published dynamical modelling. We stress that IMF constraints should be corroborated by further spectral coverage and independent methods on a galaxy-by-galaxy basis.
\end{abstract}


\section{Introduction}

The formation of the stellar content in early type galaxies (ETGs), quantified by the initial mass function (IMF), has recently been the focus of much study. The IMF sets the mass distribution of stars and affects the total mass-to-light ratio of a galaxy, as well as feedback scales from supernovae and chemical enrichment of the interstellar medium. The most direct method to constrain the IMF is through star counts, which makes measuring the IMF an immense challenge for any galaxy other than our own. The IMF in the Milky Way has been constrained by several studies, notably \citet{Salpeter1955}, who devised the classical power law form of the IMF, $\xi (m) = km^{-x}$, where the exponent takes a value of $x=2.35$, down to masses of $\sim 1\,\rmn{M_{\odot}}$. Later studies by \citet{MillerScalo1979, Kroupa2001} and \citet{Chabrier2003} incorporated a turn-off at masses below $\sim1\,\rmn{M_{\odot}}$. Constraining the low-mass end of the IMF is important as these stars dominate the mass budget of galaxies, so changes in quantity can alter the galaxy mass-to-light ($M/L$) ratio significantly.

Historically, the form of the IMF has generally been assumed to be universal between galaxies. However, recent studies have compounded the notion that the low-mass end of the IMF in fact varies as a function of galaxy properties. Several independent methods have been employed to undertake this work. Dynamical modelling of galaxy kinematics through integral field spectroscopic observations have shown that the mass `normalisation' (the $M/L$) of galaxies systematically increases with galaxy velocity dispersion \citep{Cappellari2012, Cappellari2013b}. Strong gravitational lensing results have also suggested M/L variation in ETGs \citep[e.g.][]{Treu2010, Smith2015}. However, both of these methods only probe the {\it total} mass and so suffer degeneracies between dark and luminous matter components.

In the absence of direct star counts, the most direct method to probe the stellar content of ETGs is to analyse their integrated stellar spectra. Several gravity-sensitive stellar absorption features are well documented, which vary in strength between dwarf and giant stars. The three key features in the far red region are sodium NaI $\lambda$8190 (\citealt{FaberFrench1980}, \citealt{Schiavon1997a}); the calcium triplet CaT (\citealt{Cenarro2001a, Cenarro2003}); and the Wing-Ford band FeH $\lambda$9916 (\citealt{WingFord1969}, \citealt{Carter1986}, \citealt{Schiavon1997b}, \citealt{Cushing2003}). IMF-sensitive features in the optical have also recently been utilised, including titanium oxide TiO features \citep{Spiniello2012, Ferreras2013, LaBarbera2013} and calcium monohydride CaH \citep{Spiniello2014}.

Alongside improvements in instrumentation that have allowed the detection of faint far red features, advances in stellar population modelling over recent years have also majorly contributed to our understanding of the IMF \citep[see e.g.][]{Conroy2013}. Stellar population synthesis (SPS) models now produce simple stellar population (SSP) spectra for varying IMFs \citep[e.g.][]{Vazdekis2012, ConroyVanDokkum2012a}, as well as individual elemental abundance variations \citep{ConroyVanDokkum2012a}.

Stellar population gradients within galaxies are now being routinely investigated, greatly aided by advances in integral field spectroscopy \citep[e.g.][]{Kuntschner2010, Loubser2012, Greene2015, Oliva-Altamirano2015}. Recent work by several groups has explored the possibility of radial gradients of IMF-sensitive spectral features in galaxies. Results by \citet{Martin-Navarro2015a} have indicated a gradient to the IMF slope in a massive ETG with a large central velocity dispersion, ranging from very bottom-heavy in the central region to around Salpeter further out. \citet{Martin-Navarro2015c} have also presented results indicating a massive galaxy with a constant bottom-heavy IMF independent of radius.

While these results are important and can be interpreted as a larger fraction of low-mass stars in the central, densest regions of the most massive galaxies, they are based on absorption features which are susceptible to influence from non-IMF factors. In particular they do not utilise the FeH feature, which was used by \citet{VanDokkumConroy2010, VanDokkumConroy2012, ConroyVanDokkum2012b} in their studies that reignited this field of research. \citet{Zieleniewski2015} and \citet{McConnell2016} have both presented radial measurements of NaI, CaT and FeH in several galaxies: M31 and M32 in \citeauthor{Zieleniewski2015} and two massive ETGs in \citeauthor{McConnell2016}. Both sets of results showed strong gradients in NaI but flat profiles (or slightly negative gradients in the case of the ETGs) in FeH. These resulted in discrepant predictions of the IMF between the individual features and were interpreted as normal IMFs with strong gradients in Na abundance. \citet{LaBarbera2016} have also measured radial measurements of FeH and optical TiO features in a single massive ETG. They also found a flat FeH profile contrasted with strong TiO gradients and excluded a bottom-heavy $x=3$ IMF, instead invoking a bimodal form with a taper below $1\,{\rm M}_{\odot}$.

In this paper we add to the small but growing amount of literature regarding radial gradients in IMF-sensitive stellar absorption features of massive ETGs, by presenting results from four galaxies in the Coma cluster. We target the two massive, slow-rotator, brightest cluster galaxies (BCGs) within the main cluster: NGC4889 and NGC4874 ($\epsilon=0.36, \rm{\lambda_R}=0.04$ and $\epsilon=0.12, \rm{\lambda_R}=0.08$ respectively, \citealt{Houghton2013}. $\epsilon$ is the ellipticity and $\rm{\lambda_R}$ is the specific angular momentum; slow rotators are defined as having $\rm{\lambda_R} < 0.31\sqrt{\epsilon}$, see \citealt{Emsellem2011}); as well as the BCG NGC4839 within the Coma south-west cluster. We also obtain unresolved spectroscopy for the fast rotator NGC4873 ($\epsilon=0.23, \rm{\lambda_R}=0.41$) as a low-$\sigma_*$ comparison galaxy. Our sample covers a wide range of central velocity dispersions and metallicities \citep[e.g.][]{Trager2008, Loubser2009}. Each of these properties have been purported as possible mechanisms for IMF variations (e.g. [$\alpha$/Fe]: \citealt{ConroyVanDokkum2012b}; $\sigma_*$: \citealt{ConroyVanDokkum2012b, Ferreras2013, LaBarbera2013, Spiniello2014}; [Z/H]: \citealt{Martin-Navarro2015a}), so our sample covers a suitable range to probe these correlations, and in particular a wide range of $\sigma_*$, which has been the most widely reported correlation. The IMF of Coma galaxies has been previously studied by \citet{Smith2012} using spectra stacked by velocity dispersion. However, their results only concerned the central 0.6 kpc. In this study for each BCG we present radial measurements of the far red features covered by the SWIFT instrument out to 10 arcsec (5 kpc). We then utilise the latest stellar population synthesis (SPS) models to analyse our results in the context of a possible variable IMF. 

This paper is organised as follows: Section 2 details our observations and data reduction procedures. In Section 3 we present the absorption feature strengths, and Section 4 contains our analysis using index-index diagrams and comparing to SPS model predictions for SSP spectra. We discuss our results in Section 5 and conclude our work in Section 6. In this paper we adopt the IMF naming convention used by \citet[][hereafter CvD12]{ConroyVanDokkum2012a}, namely a Chabrier IMF corresponding to that defined in \citet{Chabrier2003} for the disc of the Milky Way; a \citet{Salpeter1955} IMF defined as $x=2.35$; and a bottom-light IMF as defined in \citet{VanDokkum2008}. We also refer to the \citet{Kroupa2001} and \citet{Chabrier2003} IMFs, which are very similar in form, interchangeably as `Milky-Way' IMFs. We adopt a flat $\rmn{\Lambda}$CDM cosmology with $H_0 = 68\,\rmn{km}\,\rmn{s}^{-1}\,\rmn{Mpc}^{-1}$, $\Omega_m = 0.3$ and $\Omega_{\Lambda} = 0.7$. This gives a distance to the Coma cluster \citep[$z=0.024$,][]{HanMould1992} of $108\,\rmn{Mpc}$ and an angular scale of $0.5\,{\rm arcsec}\,{\rm kpc}^{-1}$.

\section{Observations and data reduction}

Observations were obtained over the nights of 4th May 2009; 11th May 2012; 17th, 20th and 22nd April 2013 using the SWIFT instrument \citep{Thatte2006} on the Palomar $200$ inch (5.1 m) telescope. Observations were taken at the $235$ mas spaxel$^{-1}$ spatial scale covering a field of $10'' \times 20''$. The spectra cover the wavelength range $6300-10400$\AA\ with a dispersion of $\sim 2$\AA\ FWHM and a sampling of 1\AA\ pix$^{-1}$. Table~\ref{tab:galaxy_sample} lists our galaxy sample and total exposure time for each galaxy. The seeing was around $1.5''$ for all observations.

\begin{table}
\centering
  \caption{Our sample of Coma galaxies with positions, effective radii $R_{\rm{e}}$ taken from \citet{Loubser2008} and  \citet{Houghton2013}, and total on-source exposure times. The $R_{\rm{e}}$ column shows the fraction of $R_{\rm{e}}$ covered by our observations in parentheses.}
  \label{tab:galaxy_sample}
  \begin{tabular}{@{}lcccc@{}}
  \hline
Galaxy & RA & DEC & $R_{\rm{e}}$ ($''$) & $T_{\rmn{exp}}$ (s)\\
 \hline
 NGC4889 & 13:00:08.1 & $+$27:58:37 & 38.0 (0.2) & 5400\\
 NGC4874 & 12:59:35.7 & $+$27:57:33 & 50.4 (0.2) & 6300\\
 NGC4839 & 12:57:24.3 & $+$27:29:52 & 17.2 (0.5) & 3600\\
 NGC4873 & 12:59:32.8 & $+$27:59:01 & 5.9 (0.5) & 2700\\
\hline
\end{tabular}
\end{table}

\subsection{Data reduction}

The data were reduced using the SWIFT data reduction pipeline, written in {\sc iraf} (Houghton, in preparation). The pipeline handles all the standard reduction processes of bias subtraction, flat fielding, wavelength calibration, error propagation as well as IFS specific features of illumination correction and data cube reconstruction. Cosmic rays were detected and removed using the {\sc lacosmic} routine \citep{VanDokkum2001}. First-order sky subtraction was performed by subtracting sky data cubes observed adjacent in time to each science cube. We were careful to choose sky fields as to avoid contamination from other galaxies in the cluster.

Residual sky lines and telluric absorption are two effects that significantly hamper accurate measurement of the faint far red absorption features of interest, especially the FeH index. The next two sub-sections detail our efforts to minimise these effects and ensure high quality spectra with which to make robust index measurements.

\subsection{Telluric correction}

\begin{figure*}
 \centering
 \includegraphics[width=14cm]{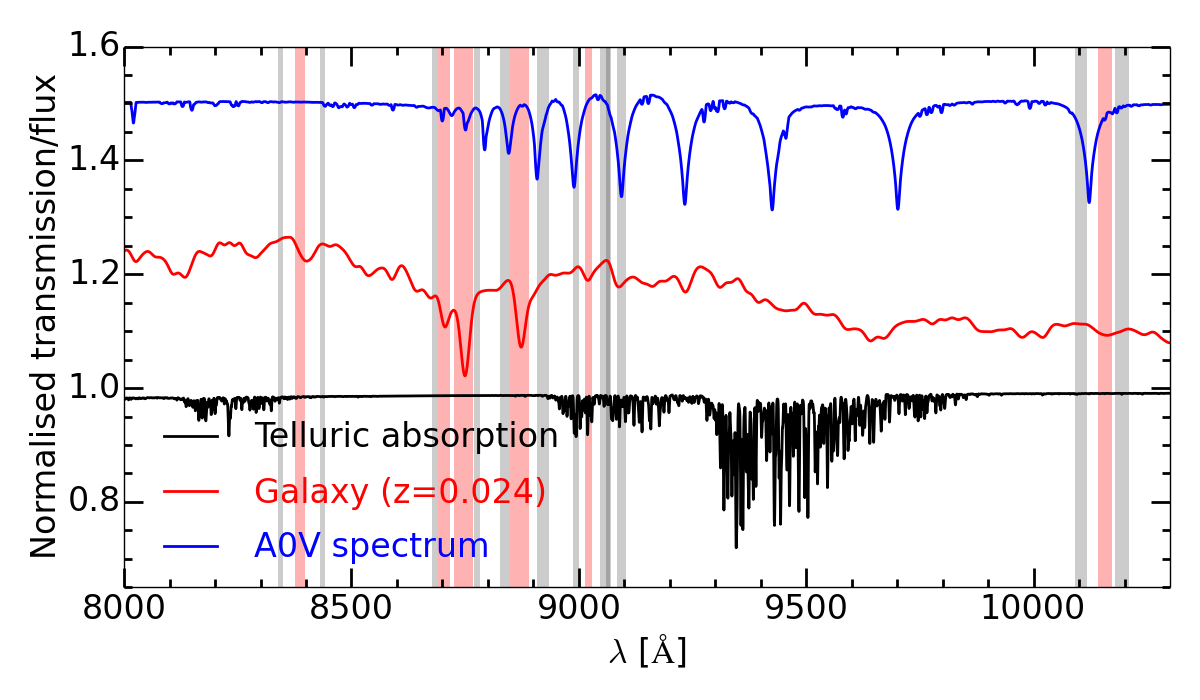}
 \caption{Plot showing atmospheric absorption spectrum (black; scaled by 0.5 and plotted at 0.98), SSP model spectrum (red; plotted at 1.2) redshifted to the nominal redshift of the Coma cluster $z=0.024$ \citep{HanMould1992}, and a model A0V spectrum (blue; continuum divided and plotted at 1.5). The telluric spectrum is convolved to the SWIFT resolution and the `galaxy' spectrum is a CvD12 13.5 Gyr, Chabrier IMF SSP that has been smoothed to $\sigma=300\,\rmn{km}\,\rmn{s}^{-1}$. The redshifted (to $z=0.024$) positions of the key absorption feature bandpasses are denoted by the shaded regions. They are, from left to right, NaI$_{\rmn{SDSS}}$, CaT, MgI, TiO and FeH.}
 \label{fig:telluric_and_galaxy_specs}
\end{figure*}

\begin{figure}
 \centering
 \includegraphics[width=8cm]{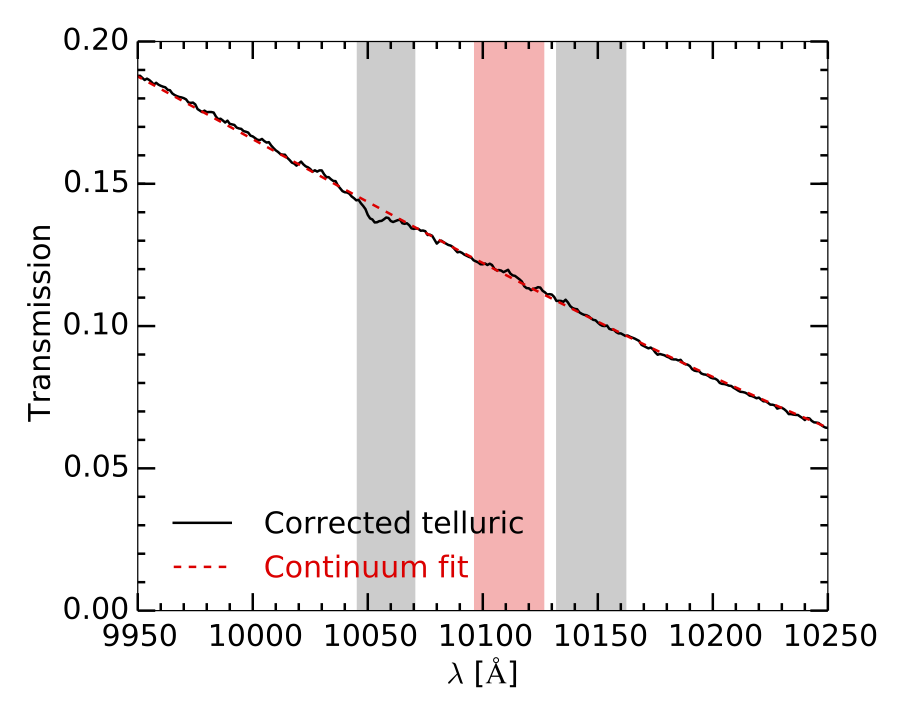}
  \caption{Plot showing the throughput, derived from a telluric spectrum, around the FeH feature after removing A0V features (black solid line) and a polynomial fit to the continuum of the telluric (red dashed line). A residual Paschen absorption feature is clearly evident at $10050$\AA\, which is left over from dividing out the A0V absorption. The shaded regions show the FeH pseudo-continuum and feature bandpass definitions redshifted to $z=0.0193$ of NGC4873 \citep{Trager2008}. The blue continuum would be affected by residuals in the telluric spectrum, but these are overcome by using the continuum fit. The gradient in the curve represents the varying throughput of SWIFT at the red end.}
 \label{fig:tellfit}
\end{figure}

Due to the redshift of the Coma cluster \citep[$z=0.024$;][]{HanMould1992} the IMF-sensitive absorption features are shifted relative to the regions affected by telluric absorption. Fig.~\ref{fig:telluric_and_galaxy_specs} shows an example telluric absorption spectrum (from the ESO Skycalc tool; \citealt{Noll2012}) convolved to the SWIFT spectral resolution and scaled by a factor of 0.5. Also plotted is an example model spectrum (red: CvD12 13.5 Gyr Chabrier IMF SSP) redshifted to $z=0.024$, as well as an example telluric standard A0V spectrum (blue: after division by a polynomial fit to the continuum) showing the positions of the prominent Paschen absorption lines. This illustrates the regions affected by both atmospheric absorption and telluric star features, relative to the positions of the key far red absorption features in the science spectra (shaded regions). Atmospheric telluric absorption is prominent in the blue pseudo-continuum of NaI and across the MgI and TiO features. The furthest red pseudo-continuum region of CaT also falls into a telluric region, but FeH is in a region of negligible telluric absorption. However the SWIFT throughput is a strong function of wavelength at the red end, which must be corrected for. Furthermore, residuals from correcting the prominent Paschen feature are present around $1.01\,\rmn{\mu m}$, which can affect our measurements of the FeH feature. Fig.~\ref{fig:tellfit} shows an example telluric spectrum along with a fit to the continuum. Also shown are the FeH bandpass definitions shifted to the redshift of NGC4873 \citep[$z=0.0193$,][]{Trager2008}. This clearly shows that our FeH measurements would be susceptible to residuals from poor fitting of the Paschen lines. We therefore pursue two separate methods to correct for telluric absorption.

Firstly, we use A0V stars observed alongside the galaxies to act as telluric standards. We remove the prominent Paschen absorption lines using a dedicated routine written in {\sc idl}. The telluric spectra are divided out by model A0V spectra provided by R. Kurucz\footnote{\url{http://kurucz.harvard.edu/stars/astar/}}, using a version of the amoeba algorithm allowing for velocity shifts and stellar rotation. Fig.~\ref{fig:allgalstelluricsky} shows examples of our telluric correction around the MgI feature for NGC4889 ($\sim$ best case), NGC4874 and NGC4839 ($\sim$ worst case). In each plot we use our kinematic template fits (see section~\ref{sec:secondsky}) as an assumed `telluric free' galaxy spectrum. We compute the rms residuals after telluric correction around each feature (NaI$_{\rmn{SDSS}}$, CaT, MgI, TiO) for each galaxy. We find that MgI has the highest residuals of $\sigma\sim0.01$ (see Fig.~\ref{fig:allgalstelluricsky}) and all other features have residuals lower than 1 per cent.

Secondly, we fit a high order polynomial to the continuum of each telluric star spectrum (after division by the best fitting A0V model) to represent a relative throughput curve without residual Paschen absorption. We then divide the galaxy by this fit to correct for the smooth throughput variation at the red end. Thus we create two separate spectra for each galaxy, one corrected by the telluric spectrum and one corrected by a continuum fit to the (A0V divided) telluric spectrum. We use the telluric corrected spectrum to measure the NaI$_{\rmn{SDSS}}$, CaT, MgI and TiO features, and we use the `throughput-divided' spectrum to measure the FeH feature. We compare FeH measurements from both methods in Fig.~\ref{fig:allFeH} in appendix~\ref{sec:fehsys}.

\subsection{Second-order sky subtraction, binning and kinematics}
\label{sec:secondsky}

Residual sky emission is still prominent at red wavelengths (after the first-order sky subtraction) due to time-dependent variations of the sky emission on the scale of minutes. These are the dominant source of systematic error around $1\,\rmn{\mu m}$ and so must be accurately removed to measure the FeH feature of only a few per cent strength relative to the continuum. We employ two independent methods to try and remove these residual lines. The first method involves fitting and removing the skylines while fitting the kinematics, as presented in \citet{Zieleniewski2015} for M31 and M32. The second method involves fitting each wavelength channel (narrowband image) of the data cube with a S\'{e}rsic function to separate the galaxy light from sky light.

\subsubsection{Simultaneous kinematics and sky subtraction}

For the first method we perform a simultaneous second-order sky subtraction while extracting the kinematics \citep{Weijmans2009}. We fit the kinematics of the spectra with the penalised pixel fitting routine \citep[{\sc ppxf}, ][]{CappellariEmsellem2004} using, as template spectra, the stellar population models from CvD12. We compute the kinematics in annular bins of increasing radii from the centres of each galaxy to achieve a target S/N $>100\,\rmn{pixel}^{-1}$ in the central part of the spectra (around CaT, MgI and TiO features). Sky subtraction was accomplished using a variation of the techniques described in \citet{Davies2007}: rather than scaling a single sky spectrum covering multiple OH vibrational transitions, we divide the spectrum into separate regions covering each OH vibrational transition. This allows independent scaling of each transition. We also scale the O$_{2}$ emission around 0.864$\,\mu$m separately. The sky spectra are extracted from a separate sky cube, which has been shifted and combined in the same way to match the spaxel positions, and from the identical annular apertures as the science spectra. All these spectra are then passed to {\sc ppxf}, which finds the best-fit linear combination (including scaling) that reproduces the skyline residuals while also finding the best fit kinematics\footnote{We manually altered the internal limits in {\sc ppxf} to allow negative sky spectrum weights.}. Furthermore, to account for flexure (error in the wavelength calibration) we shift each sky spectrum forward and backwards by three pixels when performing the fitting. For each galaxy spectrum, we measure the intrinsic instrument dispersion from the skylines at the same wavelength as the principal galaxy absorption features (NaI/CaT/FeH) and independently for each azimuthal bin.

The resolution of a galaxy spectrum as observed by SWIFT is $\sigma_{*{\rm,obs}}$ and is given by
\begin{equation}
\sigma_{*{\rm,obs}} = \sqrt{\sigma_*^2 + \sigma_{{\rm inst}}^2},
\end{equation}
where $\sigma_*$ is the intrinsic galaxy stellar velocity dispersion and $\sigma_{{\rm inst}}$ is the instrument resolution.
{\sc ppxf} measures the galaxy resolution by fitting with template spectra of intrinsic resolution $\sigma_{{\rm temp}}$. When the template resolution used in {\sc ppxf} is finer than the instrument resolution ($\sigma_{{\rm temp}}<\sigma_{{\rm inst}}$), we convolve the template library to the same resolution as the instrument, so $\sigma_{{\rm temp,n}}=\sigma_{{\rm inst}}$, and then {\sc ppxf} measures,
\begin{equation}
\sigma_{{\rm ppxf}}\approx\sigma_* = \sqrt{\sigma_{*{\rm,obs}}^2 - \sigma_{{\rm temp,n}}^2}.
\end{equation}
However, the SWIFT spectral resolution varies from $40\,\rm{km}\,\rm{s}^{-1}$ (${\rm R}\sim3100$) to $65\,\rm{km}\,\rm{s}^{-1}$ (${\rm R}\sim2000$) across the field of view and the resolution of the CvD12 models is R=2000 beyond 0.75$\,\mu$m. In this case we cannot convolve the template library to the instrument resolution and {\sc ppxf} instead underestimates the galaxy dispersion by measuring,
\begin{equation}
\sigma_{{\rm ppxf}}' = \sqrt{\sigma_{*{\rm,obs}}^2 - \sigma_{{\rm temp}}^2}.
\end{equation}
Therefore, when $\sigma_{{\rm temp}}>\sigma_{{\rm inst}}$, it is necessary to correct the dispersions found by {\sc ppxf} for the difference between instrument and template resolutions. In such cases, we added in quadrature the difference between the template resolution at that wavelength range and the instrument resolution,
\begin{equation}
\sigma_* \approx \sqrt{\sigma_{{\rm ppxf}}'^2 + (\sigma_{{\rm temp}}^2 - \sigma_{{\rm inst}}^2)}.
\end{equation}
However, in practice this only significantly affects dispersions less than $100\,\rmn{km}\,\rmn{s}^{-1}$ so has little effect with the high dispersion galaxies presented in this study.

Using this technique, we are able to clean sky residuals from the galaxy spectra (prior to calculating element abundances) by subtracting the best-fit `second-order' sky spectrum generated by {\sc ppxf}. However, the sky spectrum contains both continuum and line emission, and scaling to remove the line residuals also scales the continuum. We check whether this can be a source of bias by adding and subtracting a constant continuum level from the galaxy spectrum and checking subsequent equivalent width measurements. We use the CvD12 Chabrier and $x=3$ IMF models and measure the indices after convolving to $400\,\rmn{km}\,\rmn{s}^{-1}$ and adding or subtracting a constant level of 1, 5 and 10 per cent. Trivially and as expected we find that the index value changes by $\sim$ 1, 5 and 10 per cent respectively for both IMFs. We also find that adding a first-order gradient across a feature has minor affect, on the order of 1 per cent for a 20 per cent gradient across the feature. However, we do find that a higher-order differential change in the continuum across a feature can alter the measured index more substantially; we create a high-order polynomial fit to a 10 per cent step function offset from the central feature wavelength, and adding this changes the measured indices by around 30 per cent. As there is no reason to believe continuum and line emission scale in the same way \citep[unless the continuum is composed primarily of unresolved faint OH lines from the same transition or scattered light from bright OH lines][]{Davies2007}, we subtracted the continuum from the second-order sky spectrum before using it to remove the skyline residuals in the galaxy spectra. Typically, the effect of this in equivalent width measurements is only a few per cent, but it can be as large as 10 per cent. We use a 10$^{\rm th}$ order Legendre polynomial fit with sigma-clipping to smoothly follow the continuum and minimise bias from the sky lines. We show a comparison between our first- and second-order sky subtractions in Fig.~\ref{fig:allgalstelluricsky}, and the second-order subtraction routine is shown in more detail in Fig.~\ref{fig:allgalsfehspecs}.

For each galaxy we also obtain an {\it optimally extracted `global' spectrum} extracted over all radii. We use a method of optimal extraction based on \citet{Horne1986} and \citet{Robertson1986} to achieve maximal S/N. We analyse the optimally extracted global spectra (hereafter referred to as global spectra) along with the resolved data.

\subsubsection{Galaxy profile fitting}
\label{sec:sersic}

Due to the difficulty of removing residual skylines we pursue a separate, independent sky subtraction method to compare with the {\sc ppxf} method. In each galaxy object-minus-sky (O-S) data cube the galaxy light and sky light contribute to the total light differently; sky light $I_S$ corresponds to an additive shift up or down (assuming uniform sky across data cube), whereas galaxy light can be modelled as a S\'{e}rsic function,
\begin{equation}
I(R)=I_{\rmn{0}}e^{-kR^{\frac{1}{n}}},
\end{equation}
where $I$ is the intensity, $R$ is the radius, and $n$ is the S\'{e}rsic index \citep{Sersic1963}, and an increase in galaxy light causes a steepening of the light profile towards the centre. Thus by fitting a S\'{e}rsic profile to each wavelength channel (spatial image) of the data cube we can separate the two contributions of galaxy and sky. This profile fitting relies on the galaxy light {\it not} being uniform across the SWIFT field of view - the steeper the variation in galaxy light the better our ability to disentangle sky and galaxy. We perform this procedure for each O-S data cube. We iterate over all wavelengths and first perform a `free' fit of,
\begin{equation}
I(R)+I_S,
\end{equation}
using the amoeba algorithm and allowing all parameters to vary. Then we fit a `fixed' S\'{e}rsic to each wavelength channel, varying only $I_0$ and $I_S$ with the other variables fixed at the mean values from the `free' fit. Finally we perform another fixed fit after masking out any bad IFU slices where the fixed fit varies from the science data cube by over two standard deviations.

We are left with sky-subtracted O-S data cubes, which we combine to create an alternative sky-subtracted data cube for each galaxy. We compute the binning and kinematics using {\sc ppxf} as before but without performing second-order sky subtraction described in Section~\ref{sec:secondsky}. Thus we are left with three different spectra for each galaxy radial bin (and three different global spectra), which we summarise here:\\
(a) telluric corrected spectra sky-subtracted using {\sc ppxf},\\
(b) throughput corrected spectra sky-subtracted using {\sc ppxf},\\
(c) throughput corrected spectra sky-subtracted using O-S spatial fitting.\\
Spectra (a) are used for measuring NaI, CaT, MgI and TiO features. Spectra (b) are used for measuring the FeH feature and we compare to measurements from spectra (a) and (c) to see the effects of telluric correction and sky-subtraction methods on the FeH index.

\subsubsection{Masking}

After the telluric correction and sky-subtraction routines a small number of the resolved spectra still contain bad pixels near index features or continuum definitions. We mask these regions out and replace with values from our best fit kinematic template spectra. No masking is required for any NGC4889 or NGC4873 spectra. NGC4874 has two narrow bad pixel regions around CaT and a bad region in the blue continuum of MgI. The outermost spectrum of NGC4839 suffers from a large number of bad regions around NaI and CaT and we treat these index measurements with some caution. No masking is required for the global spectra.

\begin{table}
 \centering
  \caption{Median S/N of spectra around each index for each galaxy. Global refers to the optimally extracted spectrum extracted over all radii for each galaxy.}
  \label{tab:spec_snrs}
  \begin{tabular}{@{}lccccc@{}}
  \hline
R ($''$) & NaI$_{\rmn{SDSS}}$ & CaT &  MgI & TiO & FeH\\
 \hline
 \multicolumn{6}{c}{NGC4889} \\
 0.0 & 121 & 123 & 124 & 124 & 65 \\
 1.7 & 155 & 159 & 160 & 160 & 83 \\
 2.6 & 156 & 160 & 163 & 164 & 82 \\
 3.6 & 144 & 147 & 152 & 154 & 76 \\
 4.5 & 127 & 133 & 141 & 144 & 67 \\
 5.4 & 112 & 116 & 123 & 127 & 58 \\
 6.3 & 91 & 95 & 101 & 105 & 48 \\
 7.5 & 90 & 92 & 99 & 102 & 48 \\
 Global & 349 & 355 & 368 & 372 & 200 \\
 \multicolumn{6}{c}{NGC4874} \\
 0.0 & 66 & 67 & 70 & 70 & 33 \\
 1.8 & 96 & 99 & 103 & 104 & 49 \\
 2.8 & 94 & 96 & 102 & 104 & 48 \\
 3.7 & 93 & 96 & 102 & 104 & 47 \\
 4.8 & 90 & 94 & 100 & 103 & 45 \\
 6.8 & 100 & 105 & 114 & 118 & 51 \\
 Global & 213 & 215 & 226 & 231 & 115 \\
 \multicolumn{6}{c}{NGC4839} \\
 0.0 & 63 & 64 & 66 & 67 & 35 \\
 2.0 & 99 & 101 & 107 & 108 & 55 \\
 5.7 & 67 & 67 & 75 & 76 & 44 \\
 Global & 142 & 145 & 154 & 155 & 85 \\
 \multicolumn{6}{c}{NGC4873} \\
 Global & 96 & 95 & 105 & 105 & 54 \\
\hline
\end{tabular}
\end{table}

\subsection{Index measurements}

The S/N of our spectra around each index are shown in Table~\ref{tab:spec_snrs}. Before making index measurements, we de-redshift each spectrum to correct for its velocity as determined from the kinematic fit. 

\begin{table}
\centering
\caption{Optical and far red index bandpass and continuum definitions used in this paper. The optical definitions are taken from \citet{Worthey1994}, and the far red definitions are taken from \citet{Cenarro2001a} and \citet{ConroyVanDokkum2012a}. In this work we use the NaI$_{\rmn{SDSS}}$ index as defined in \citet{LaBarbera2013}. The TiO index is defined as the ratio between the blue and red pseudo-continua. Wavelengths are in vacuum.}
  \label{indices}
  \begin{tabular}{@{}lccc@{}}
  \hline
Index & Blue Continuum & Feature & Red Continuum\\
          & (\AA) & (\AA) & (\AA)\\
  \hline
  \multicolumn{4}{c}{{\bf Optical features}}\vspace{3pt}\\
 H$\beta$ & 4829.2-4849.2 & 4849.2-4878.0 & 4878.0-4893.0\\
 Fe52 & 5234.7-5249.7 & 5247.2-5287.2 & 5287.2-5319.7\\
  Fe53 & 5306.1-5317.4 & 5313.6-5353.6 & 5354.9-5364.9\\
 Mg{\it b} & 5144.1-5162.8 & 5161.6-5194.1 & 5192.8-5207.8\\
 \hline
 \multicolumn{4}{c}{{\bf Far red features}}\vspace{3pt}\\
 NaI$_{\rmn{SDSS}}$ & 8145.2-8155.2 & 8182.3-8202.3 & 8235.3-8246.3\\
 CaT & 8474.0-8484.0 & 8484.0-8513.0 & 8563.0-8577.0\\
         & 8474.0-8484.0 & 8522.0-8562.0 & 8563.0-8577.0\\
         & 8619.0-8642.0 & 8642.0-8682.0 & 8700.0-8725.0\\
 MgI & 8777.4-8789.4 & 8801.9-8816.9 & 8847.4-8857.4\\
  TiO & 8835.0-8855.0 &  & 8870.0-8890.0 \\
 FeH & 9855.0-9880.0 & 9905.0-9935.0 & 9940.0-9970.0\\
\hline
\end{tabular}
\end{table}

\begin{figure*}
 \centering
 \includegraphics[width=16cm]{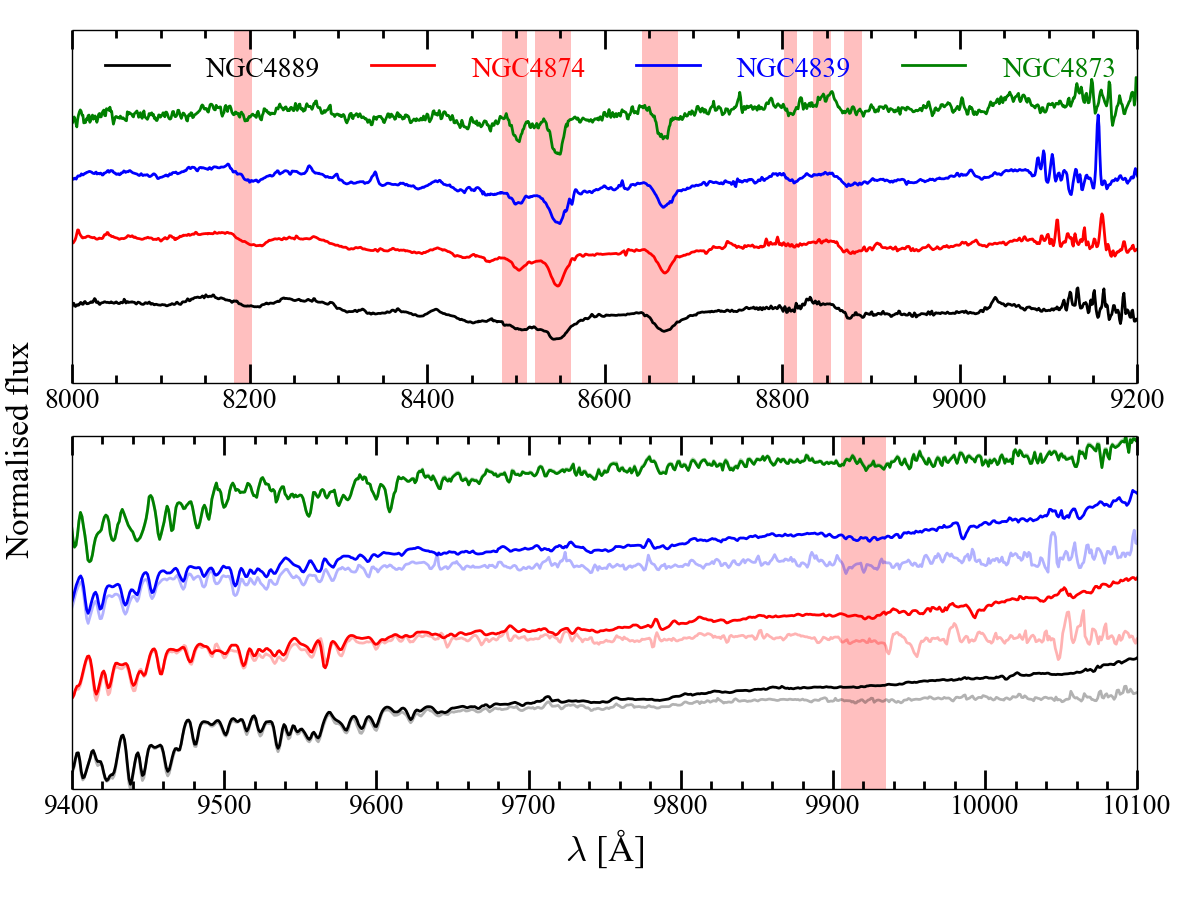}
 \caption{Global (optimally extracted) spectra of the four Coma galaxies NGC4889 (black), NGC4874 (red), NGC4839 (blue) and NGC4873 (green). Spectra have been normalised by the median continuum level and are vertically spaced for presentation purposes. In the lower panel the fainter coloured lines show the global spectra obtained using the O-S cube S\'{e}rsic profile fitting discussed in section~\ref{sec:sersic}. The spectra in the lower panel have also been divided by a continuum fit to the telluric spectra and so still show telluric features around $9100$--$9600$\AA. The red shaded regions indicate the bandpass positions for the five far red indices.}
 \label{fig:gspecs}
\end{figure*}

\begin{figure*}
\centering
\begin{subfigure}{.325\textwidth}
  \centering
  \includegraphics[width=.99\linewidth]{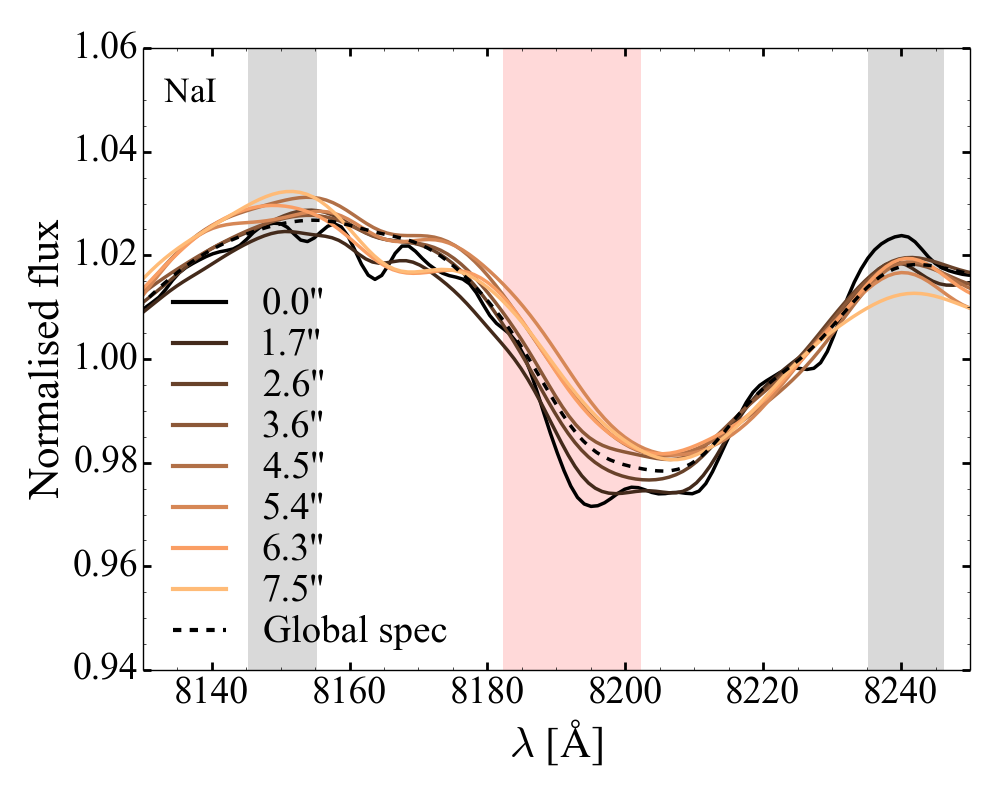}
\end{subfigure}
\begin{subfigure}{.325\textwidth}
  \centering
  \includegraphics[width=.99\linewidth]{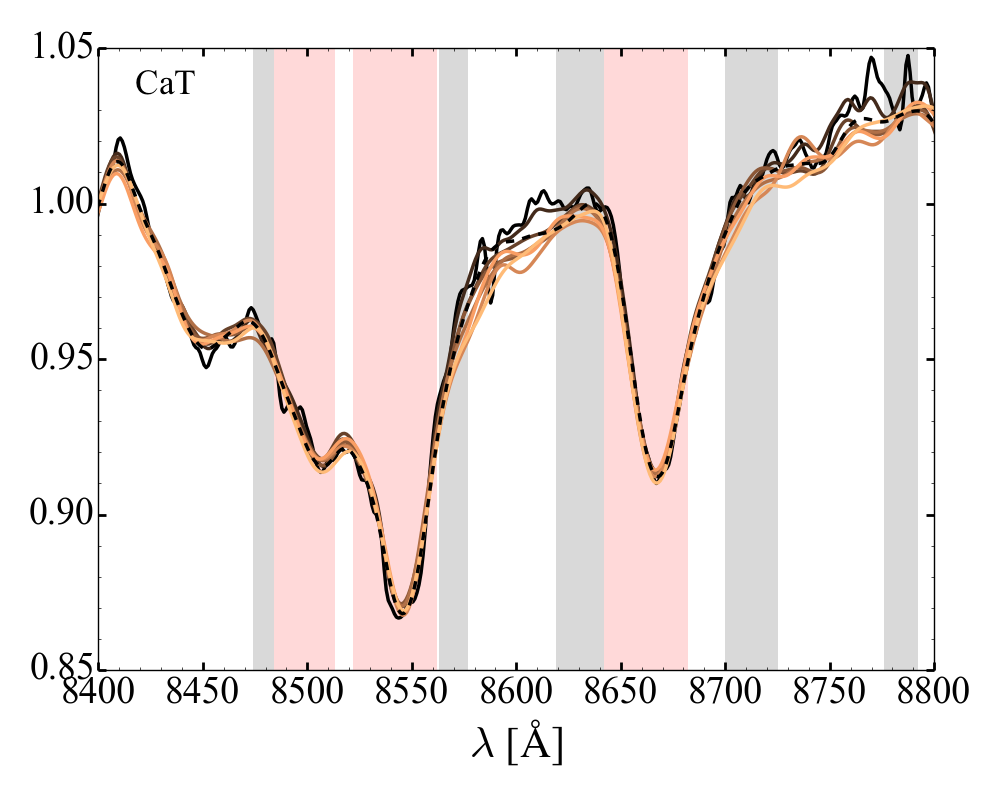}
\end{subfigure}
\begin{subfigure}{.325\textwidth}
  \centering
  \includegraphics[width=.99\linewidth]{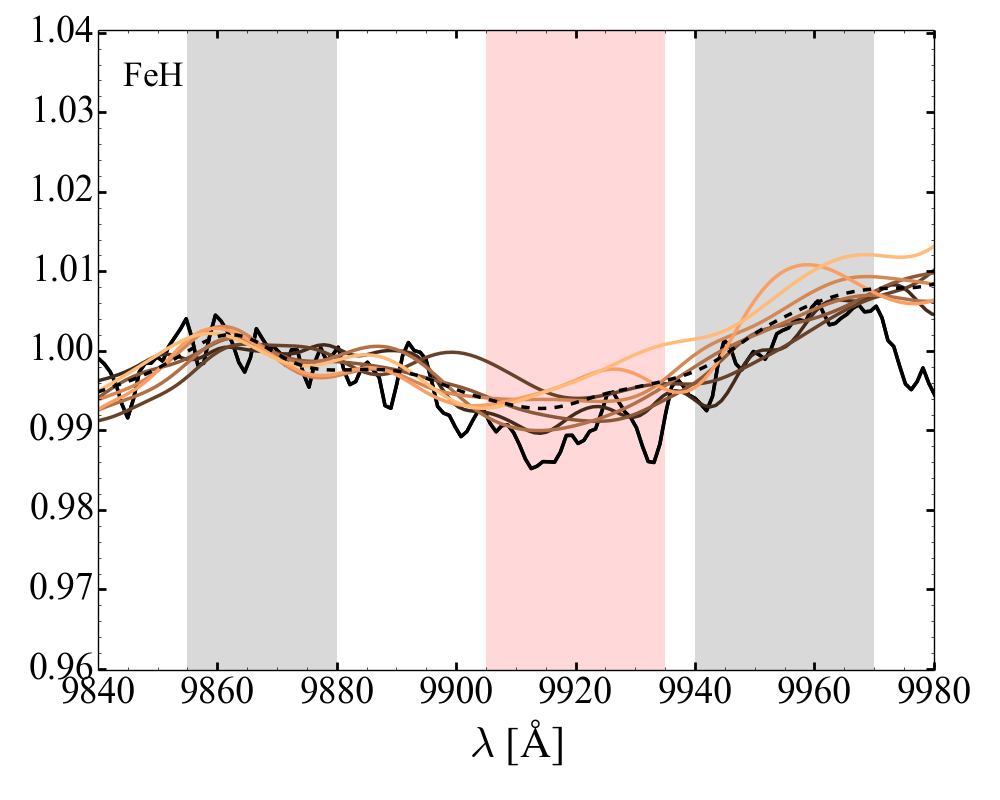}
\end{subfigure}
\caption{Spectra showing the three IMF-sensitive features in NGC4889: from left to right: NaI$_{\rmn{sdss}}$, CaT and FeH. The spectra have been convolved to a common velocity dispersion of $400\,\rmn{km}\,\rmn{s}^{-1}$. The spectra are coloured by radial distance in arcsec from the galaxy centre. Also plotted is the optimally extracted global spectrum (black dashed line). The limits of the feature and pseudo-continua band definitions are shown as the shaded red and grey regions respectively.}
\label{fig:NGC4889specs}
\end{figure*}

\begin{figure*}
\centering
\begin{subfigure}{.325\textwidth}
  \centering
  \includegraphics[width=.99\linewidth]{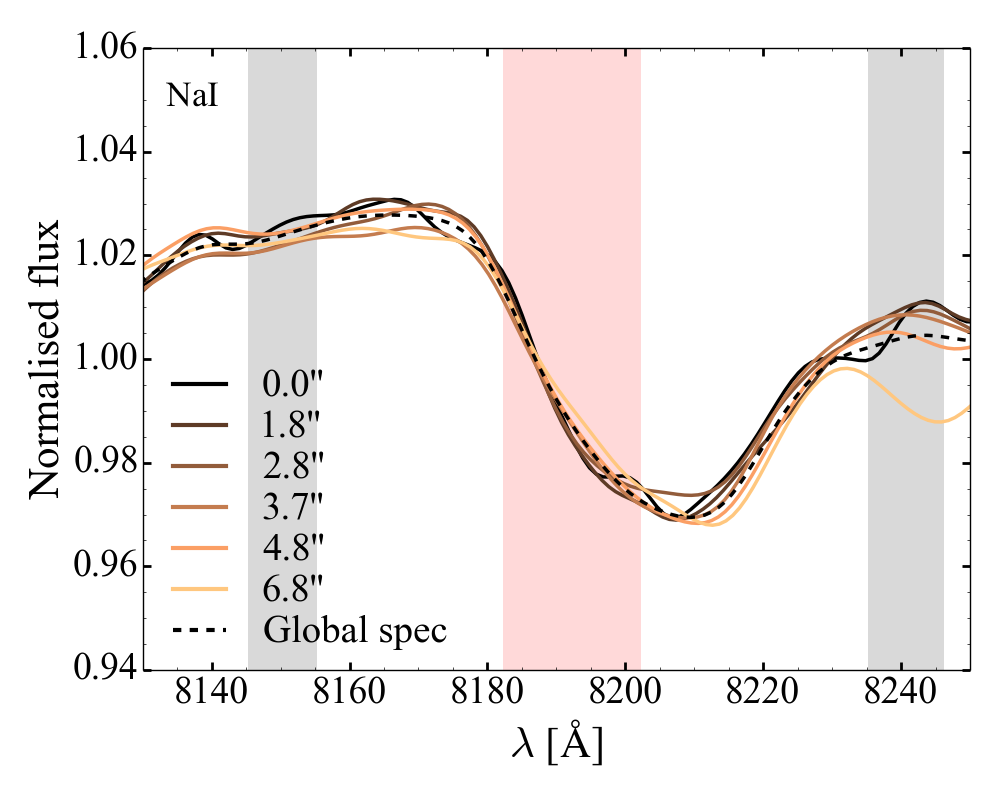}
\end{subfigure}
\begin{subfigure}{.325\textwidth}
  \centering
  \includegraphics[width=.99\linewidth]{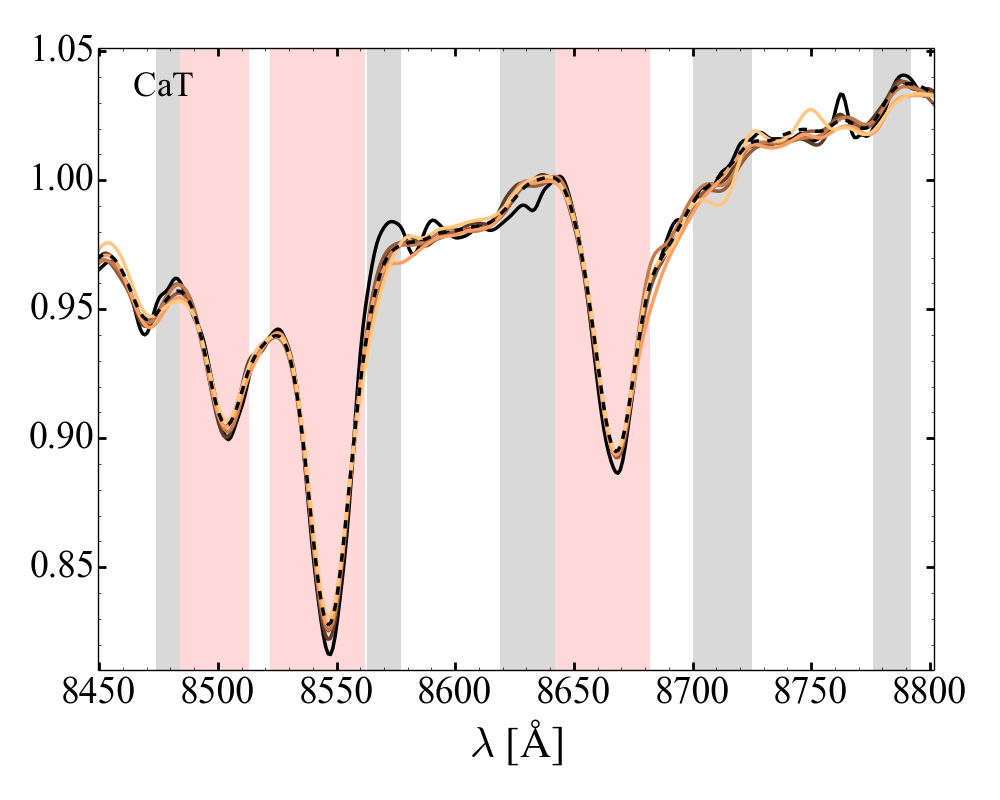}
\end{subfigure}
\begin{subfigure}{.325\textwidth}
  \centering
  \includegraphics[width=.99\linewidth]{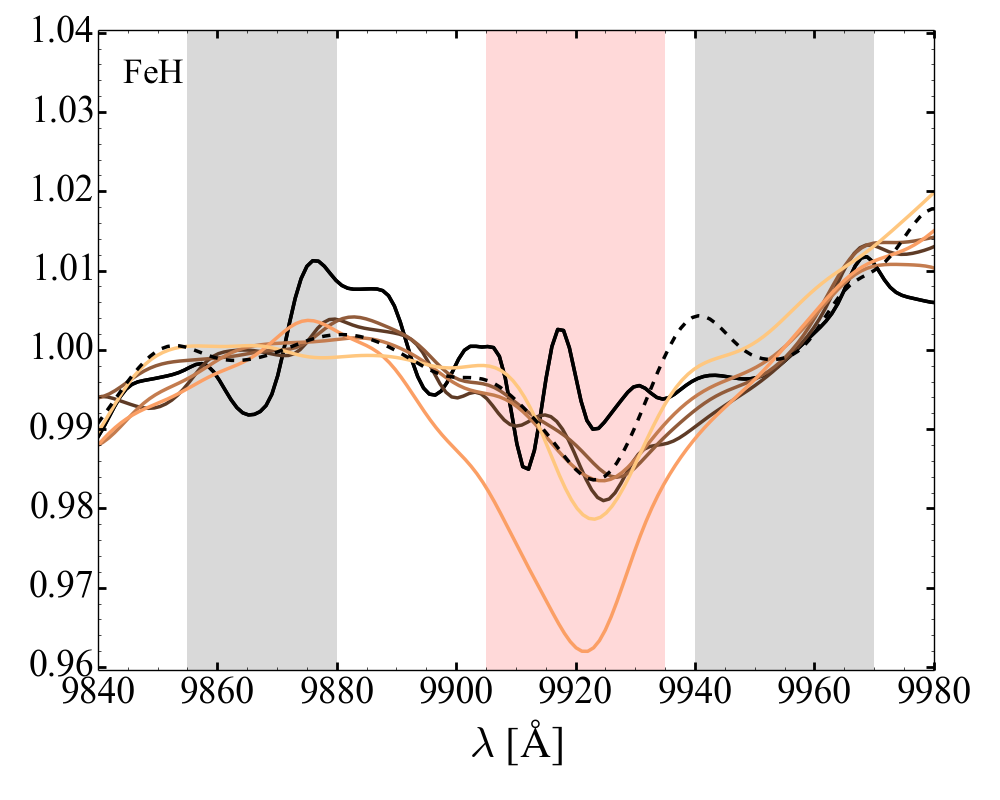}
\end{subfigure}
\caption{Same as Fig.~\ref{fig:NGC4889specs} but for NGC4874: spectra showing the three IMF-sensitive features in NGC4874: from left to right: NaI$_{\rmn{sdss}}$, CaT and FeH. These spectra are convolved to $270\,\rmn{km}\,\rmn{s}^{-1}$.}
\label{fig:NGC4874specs}
\end{figure*}

\begin{figure*}
\centering
\begin{subfigure}{.325\textwidth}
  \centering
  \includegraphics[width=.99\linewidth]{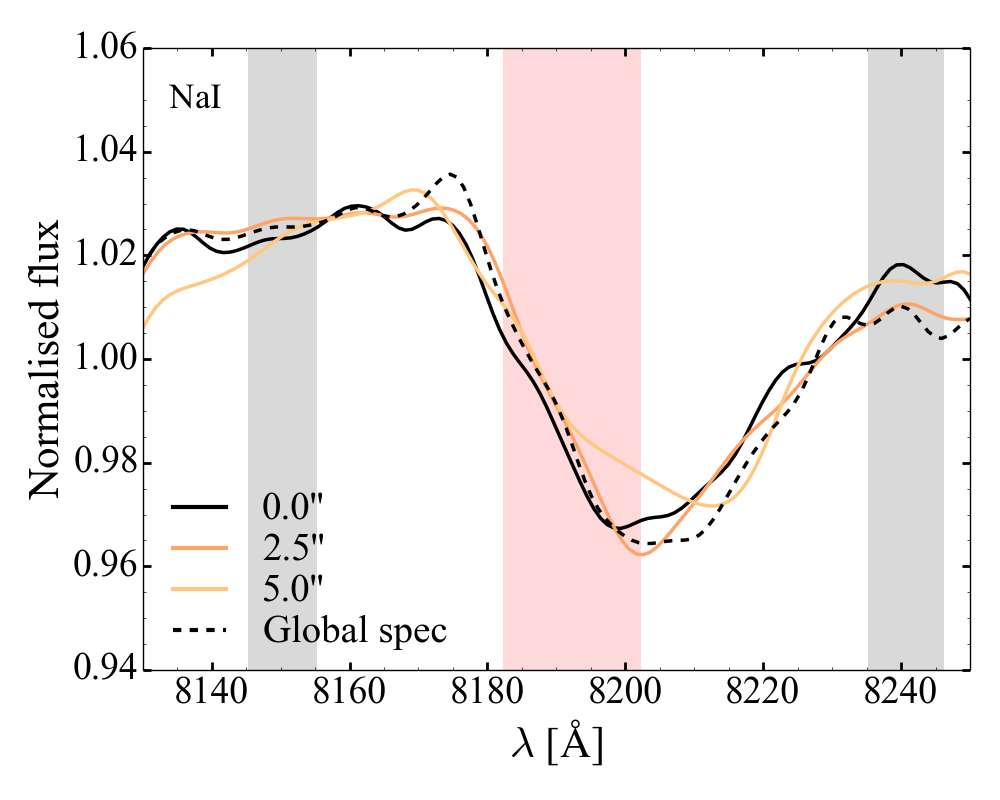}
\end{subfigure}
\begin{subfigure}{.325\textwidth}
  \centering
  \includegraphics[width=.99\linewidth]{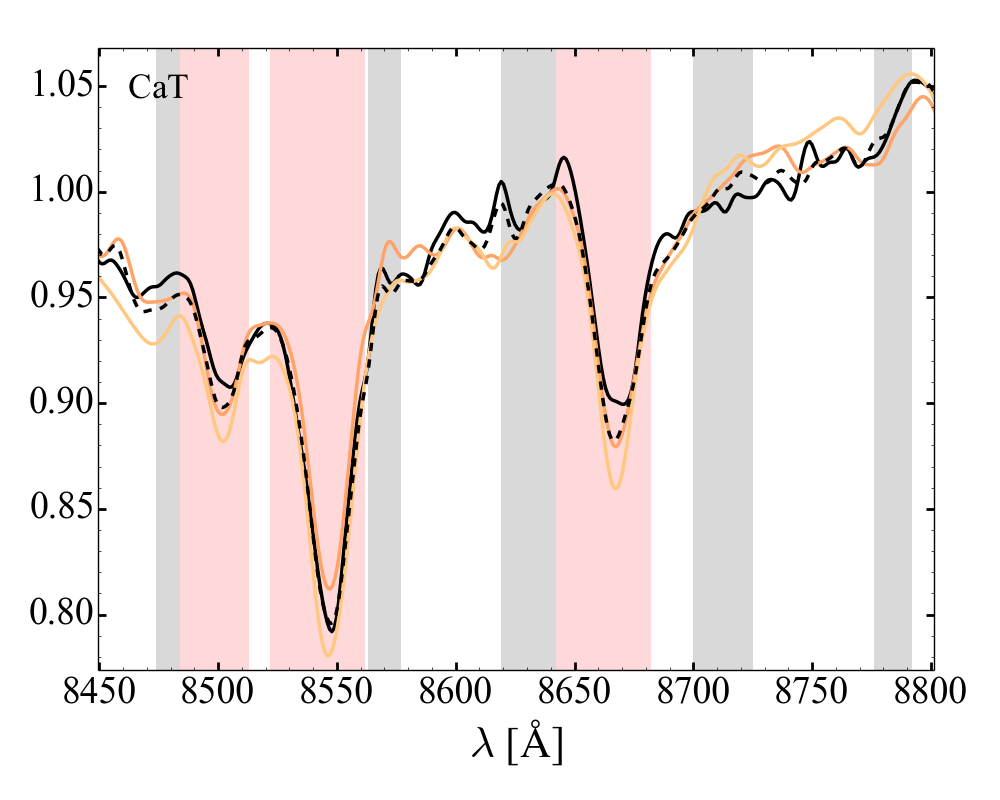}
\end{subfigure}
\begin{subfigure}{.325\textwidth}
  \centering
  \includegraphics[width=.99\linewidth]{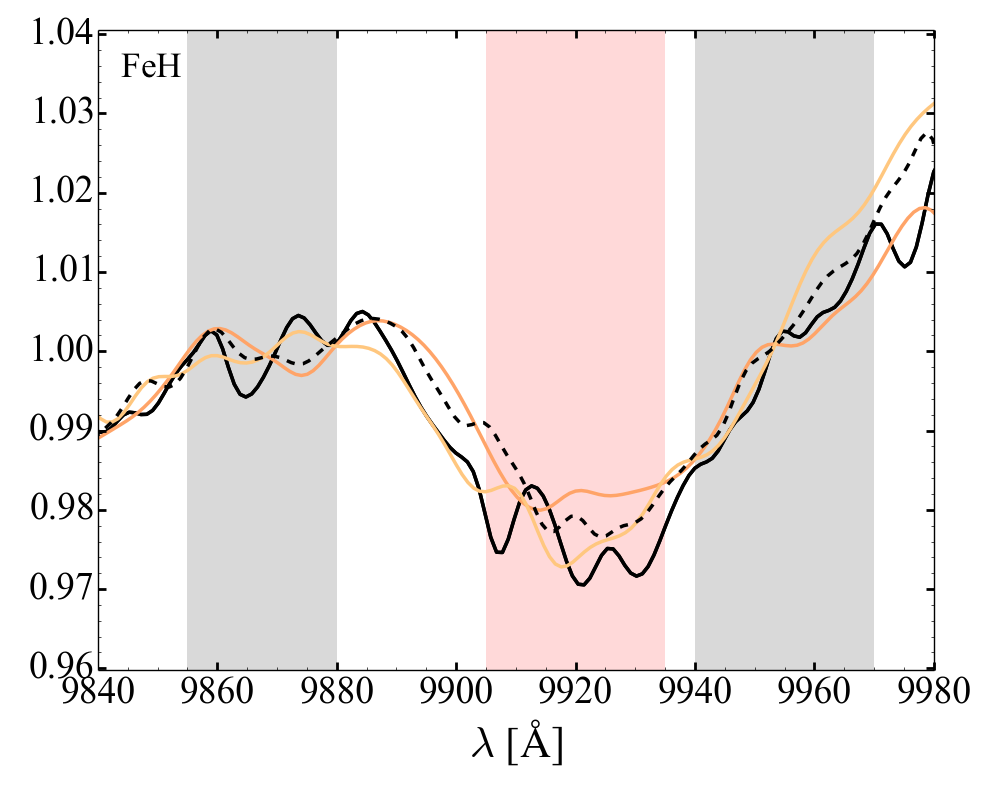}
\end{subfigure}
\caption{Same as Fig.~\ref{fig:NGC4889specs} but for NGC4839: spectra showing the three IMF-sensitive features in NGC4839: from left to right: NaI$_{\rmn{sdss}}$, CaT and FeH.  These spectra are convolved to $270\,\rmn{km}\,\rmn{s}^{-1}$.}
\label{fig:NGC4839specs}
\end{figure*}

\begin{figure*}
\centering
\begin{subfigure}{.325\textwidth}
  \centering
  \includegraphics[width=.99\linewidth]{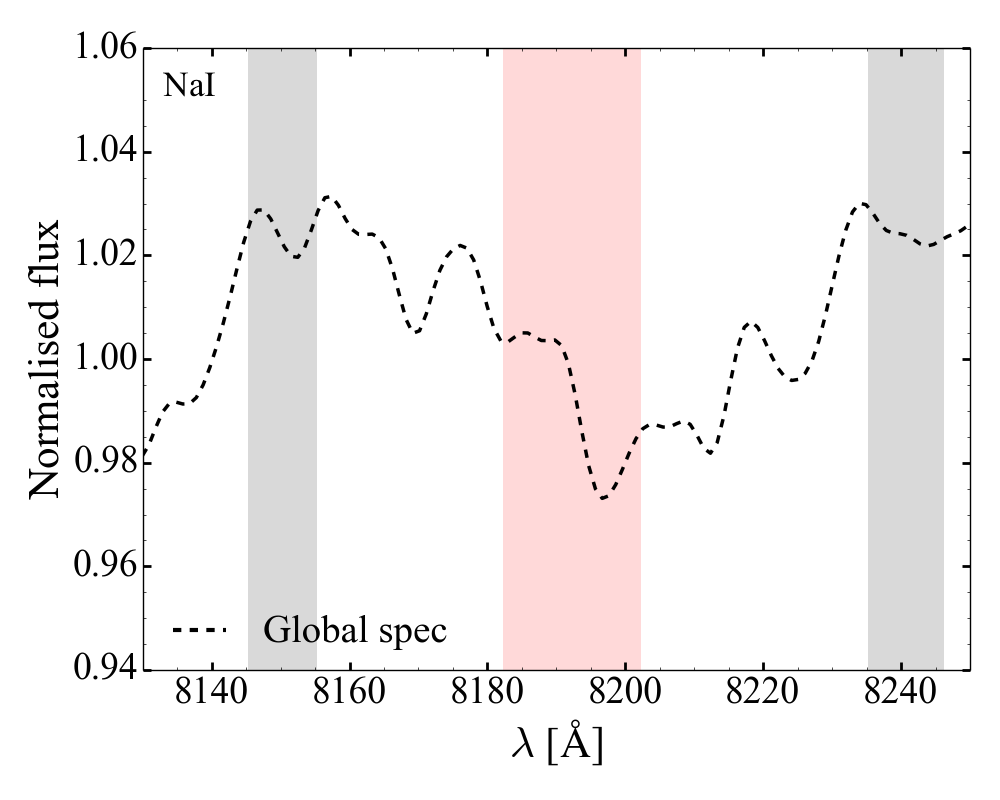}
\end{subfigure}
\begin{subfigure}{.325\textwidth}
  \centering
  \includegraphics[width=.99\linewidth]{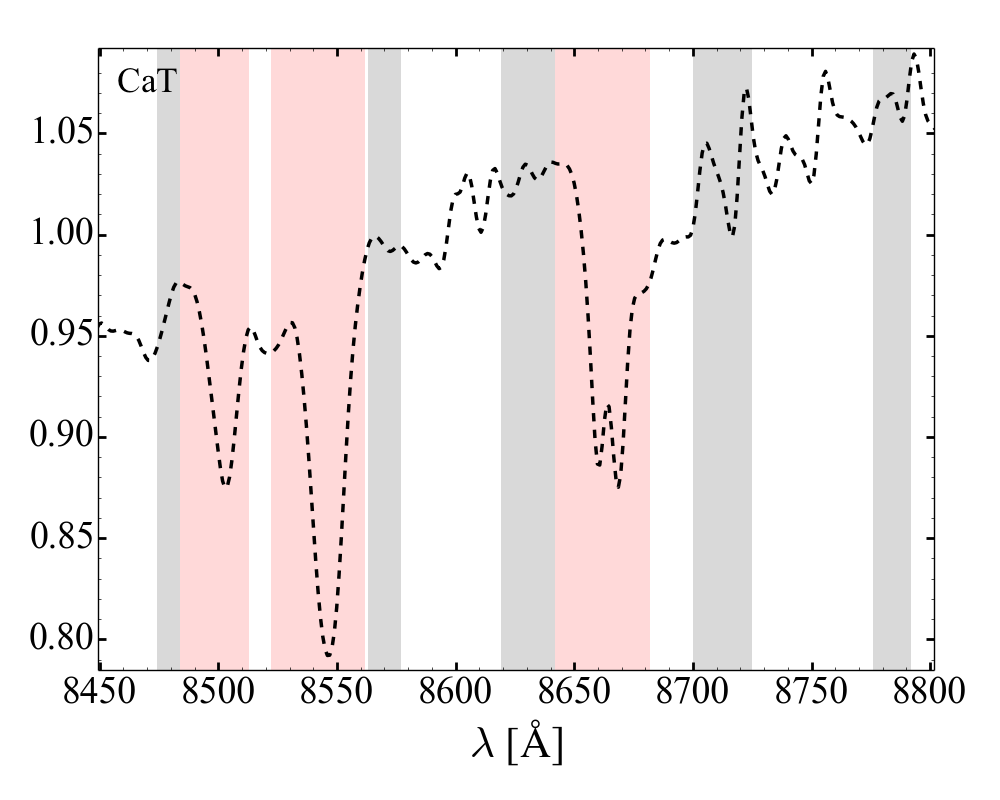}
\end{subfigure}
\begin{subfigure}{.325\textwidth}
  \centering
  \includegraphics[width=.99\linewidth]{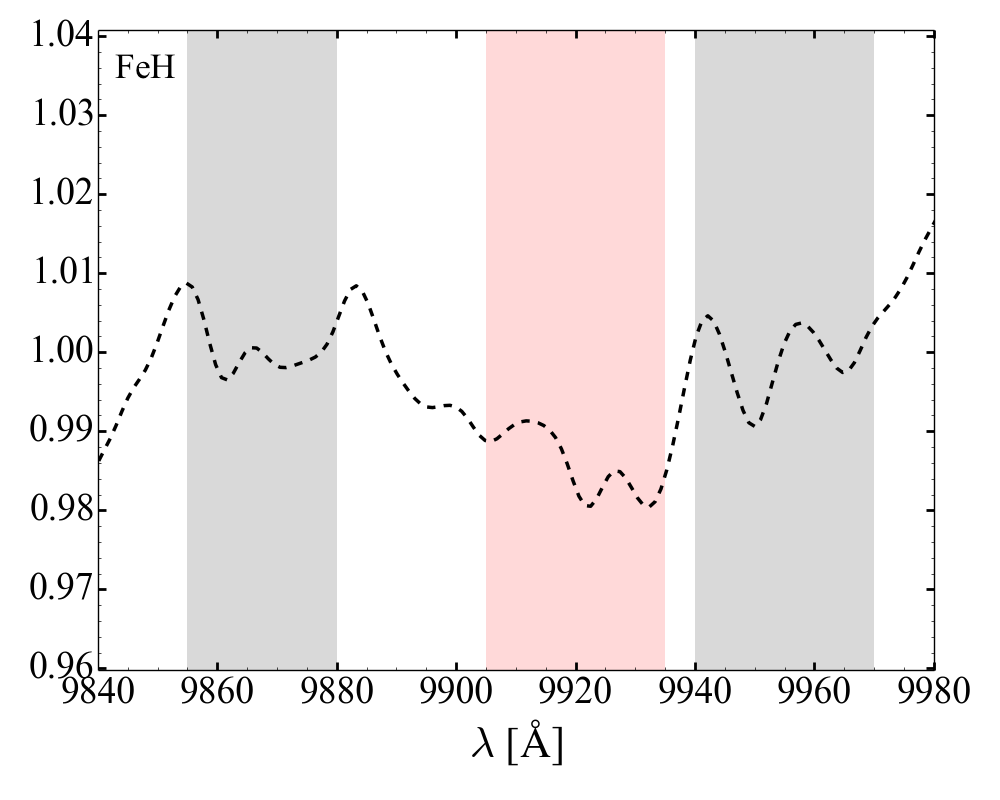}
\end{subfigure}
\caption{Same as Fig.~\ref{fig:NGC4889specs} but for NGC4873: global spectra showing the three IMF-sensitive features in NGC4873: from left to right: NaI$_{\rmn{sdss}}$, CaT and FeH.  These spectra are convolved to $185\,\rmn{km}\,\rmn{s}^{-1}$.}
\label{fig:NGC4873specs}
\end{figure*}

The global galaxy spectra are shown in Fig.~\ref{fig:gspecs}. The resolved spectra around the NaI, CaT and FeH features are plotted for each of the BCGs in Figs.~\ref{fig:NGC4889specs}, \ref{fig:NGC4874specs} and \ref{fig:NGC4839specs}, and the for global spectra of NGC4873 in Fig.~\ref{fig:NGC4873specs}.

In order to equally compare the subsequent equivalent width measurements within each galaxy, as well as between galaxies, we correct all our index measurements to a common dispersion of $200\,{\rm km}\,{\rm s}^{-1}$. We describe this procedure in Appendix~\ref{sec:corrsig}. We measure the index equivalent widths using the formalism for generic indices given in \citet{Cenarro2001a}, which includes an error-weighted least-squares fit to the pseudo-continuum. We propagate individual pixel photon errors through the data reduction pipeline as variance spectra for each science spectrum, and present our index measurements with formal $1\sigma$ uncertainties.

\subsubsection{Index measurement robustness}

To test the robustness of our results, we derive systematic uncertainties on the FeH index measurements from our multiple sky subtraction methods. The spectra and measurements are presented fully in Appendix~\ref{sec:fehsys}. We use the spread of our FeH measurements as a systematic uncertainty on our results in the following sections.

\section{results}

\begin{table*}
\centering
  \caption{Index equivalent widths from the optimally extracted global spectra for each galaxy, measured at a common velocity dispersion of $\sigma=200\,{\rm km}\,{\rm s}^{-1}$. For each BCG, for the FeH feature we present both the random uncertainty and ($^{{\rm max}}_{{\rm min}}$) values due to systematic variation from our sky subtraction procedures.}
  \label{gindices}
  \begin{tabular}{@{}lccccc@{}}
  \hline
Galaxy & NaI$_{\rmn{SDSS}}$ & CaT & MgI & TiO & FeH \\
          & (\AA) & (\AA) & (\AA) & & (\AA) \\
 \hline
 NGC4889 & $0.80\pm0.02$ & $6.87\pm0.05$ & $0.42\pm0.02$ & $1.078\pm0.001$ & $0.30\pm0.04$ ($^{0.30}_{0.19}$) \vspace{3pt}\\
 NGC4874 & $0.53\pm0.03$ & $6.58\pm0.09$ & $0.19\pm0.02$ & $1.057\pm0.001$ & $0.39\pm0.07$ ($^{0.41}_{0.33}$) \vspace{3pt}\\
 NGC4839 & $0.68\pm0.05$ & $6.92\pm0.13$ & $0.35\pm0.03$ & $1.059\pm0.002$ & $0.64\pm0.10$ ($^{0.64}_{0.20}$) \vspace{3pt}\\
 NGC4873 & $0.62\pm0.07$ & $7.65\pm0.12$ & $0.59\pm0.03$ & $1.071\pm0.002$ & $0.42\pm0.14$ \\
\hline
\end{tabular}
\end{table*}

\subsection{Global spectra}

For each galaxy we have a global spectrum optimally extracted over all radii. Fig.~\ref{fig:indicesvr} shows the index measurements from these global spectra as white-filled symbols located at 4 arcsec, along with the resolved spectra at different radii (colour filled symbols). For NGC4873 we only measure indices from the global spectra. For completeness we present the global index measurements and uncertainties in Table~\ref{gindices}, measured at a common resolution of $\sigma=200\,{\rm km}\,{\rm s}^{-1}$.

\subsection{Resolved spectra}

Radial gradients of absorption are shown in Fig.~\ref{fig:indicesvr} and we discuss the trends for each galaxy in turn. For NGC4873 our data did not have sufficient S/N to create resolved spectra, so this section covers only the three BCGs.

\subsubsection{NGC4889}

Strong negative gradients are evident for NaI$_{\rm{SDSS}}$, CaT and MgI. Sodium ranges from $\sim1.0$\AA\ in the central bin ($0.5\,\rm{kpc}$) to $\sim0.6$\AA\ at 8 arcsec ($4\,\rm{kpc}$). Calcium ranges from $7.6$\AA\ in the central bin to flat at $\sim6.5$\AA\ from 5 arcsec ($2.5\,\rm{kpc}$) outwards. Magnesium ranges from $\sim0.7$\AA\ in the central 2 arcsec ($1\,\rm{kpc}$) to $\sim0.25$\AA\ at 8 arcsec. A weak negative gradient in TiO is evident, varying from 1.081 in the central 3 arcsec ($1.5\,\rm{kpc}$) to 1.075 from 6 arcsec ($3\,\rm{kpc}$). Finally, FeH shows a flat profile throughout of around $0.30$\AA, although the random and systematic uncertainties place it anywhere between $0.1-0.5$\AA.

\subsubsection{NGC4874}

Negative gradients are evident in CaT and MgI, and flat profiles for NaI${\rm{SDSS}}$, TiO and FeH. Calcium ranges from $7.1$\AA\ in the central bin to $6.3$\AA\ at 8 arcsec. Magnesium ranges from $\sim0.6$\AA\ in the central $\sim3\,\rm{arcsec}$ to $0.4$\AA\ at 8 arcsec. Sodium absorption is flat at $\sim0.6$\AA\ throughout the central 6 arcsec with the outermost bin showing weaker absorption at $0.3$\AA, although we see from Fig.~\ref{fig:NGC4874specs} that the red continuum of the outermost spectrum has been strongly affected by residual sky over-subtraction. TiO shows similar behaviour with a flat profile at 1.060 to 5 arcsec and a slight decrease to 1.055 at $\sim7\,\rm{arcsec}$. FeH absorption scatters around $0.4$\AA\ at all radii but carries a large uncertainty. We also note one clear anomalous FeH measurement at 5 arcsec showing very strong absorption. 

\subsubsection{NGC4839}

\begin{figure*}
 \centering
 \includegraphics[width=14.6cm]{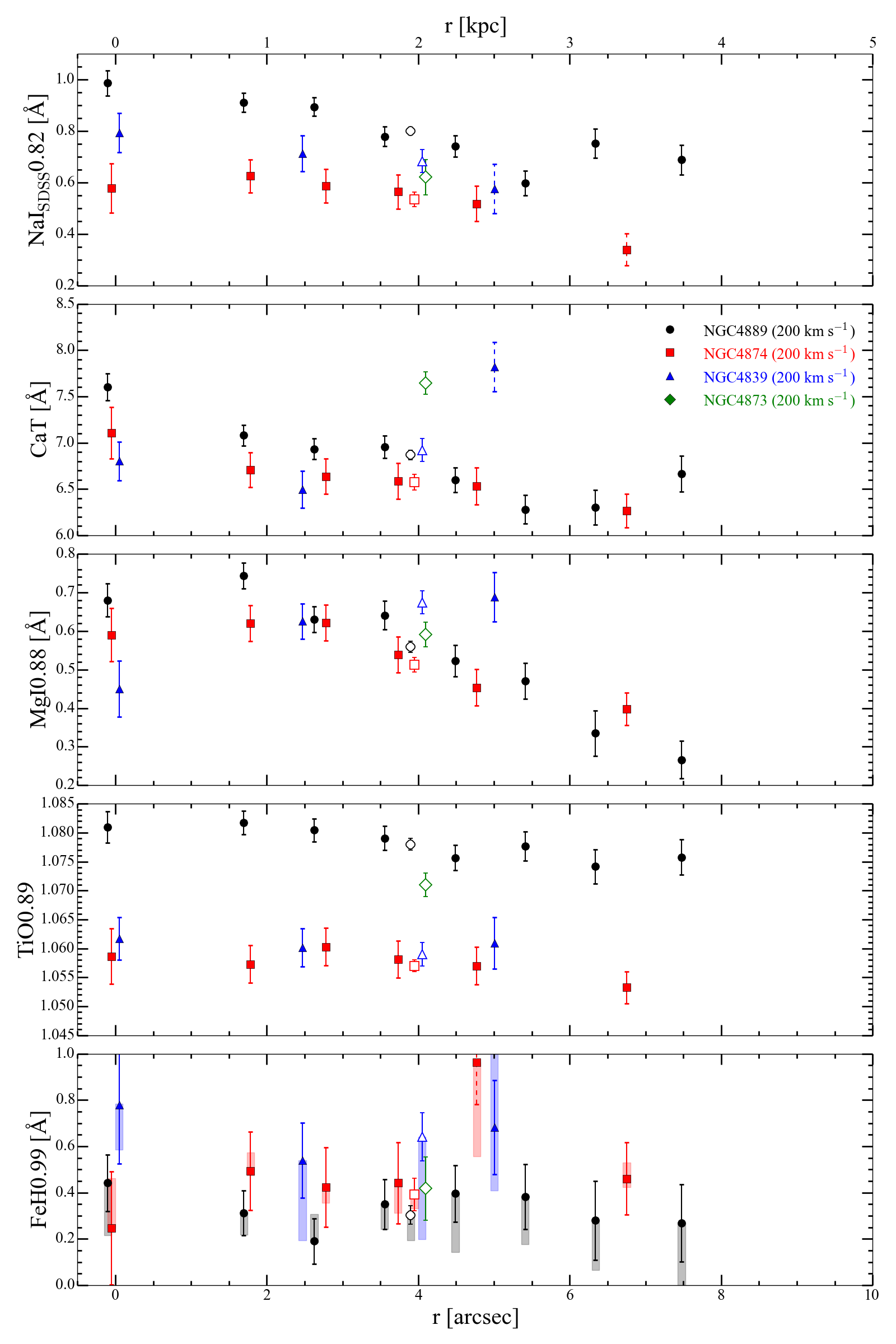}
 \caption{Plots of radial gradients for each index for NGC4889 (black circles), NGC4874 (red squares), NGC4839 (blue triangles), and NGC4873 (green diamonds). The index equivalent width measurements have been corrected to a common velocity dispersion of $200\,{\rm km}\,{\rm s}^{-1}$. Plotted in white-filled symbols at $4''$ are the values from the optimally extracted global spectra. The points with dashed error bar lines have been affected by prominent residual sky. In the FeH sub-panel the shaded rectangles show the systematic uncertainty for each BGC measurement, derived through the various sky subtraction methods detailed in Sections~\ref{sec:secondsky} and \ref{sec:sersic}. The individual FeH measurements from each reduction method are shown for each BCG in Fig.~\ref{fig:allFeH}.}
 \label{fig:indicesvr}
\end{figure*}

Due to the shorter total exposure time on this galaxy we only create three radial bins with sufficient S/N to make measurements. The outermost spectrum is also strongly affected by residual sky features, which are marked on Fig.~\ref{fig:indicesvr} with dashed error bar lines for NaI$_{\rm{SDSS}}$ and CaT. The sodium feature shows strong central absorption of $\sim0.8$\AA\ with a negative gradient to $0.7$\AA\ at 3 arcsec ($1.5\,\rm{kpc}$). Calcium displays flat absorption of $\sim6.6$\AA\ out to 3 arcsec. MgI shows absorption at $0.45$\AA\ in the central 3 arcsec and the second and third bins suggest a positive gradient up to $\sim0.65$\AA\ around 6 arcsec. TiO displays a flat profile at a similar level to NGC4874 of $\sim1.060$. FeH shows a strong central feature of $\sim0.8$\AA, lowering to a flat profile of between 0.5--0.6\AA. However, the large random and systematic uncertainties on this galaxy's FeH measurements place the measurements anywhere between 0.2--0.7\AA\ for all other points except the central value. Therefore there is possible suggestion that NGC4839 shows a stronger FeH feature compared to the two BCGs in the main Coma cluster.

\section{Analysis}

In this section we compare our measurements with SPS model predictions, in combination with previously published optical indices. We firstly use optical indices to constrain the age, metallicity [Z/H], $\alpha$-enhancement [$\alpha$/Fe] and iron-enhancement [Fe/H] of the galaxies, before using the far red indices to constrain the IMF slope.

\subsection{Stellar population synthesis models}

We make use of two sets of SPS models. For the optical features we use the \citet[][hereafter V15]{Vazdekis2015} models, which cover the wavelength range 3540.5--7409.6\AA\ for a range of IMFs, ages, metallicities, and $\alpha$-enhancements. We use the V15 models with BaSTI isochrones \citep{Pietrinferni2004, Pietrinferni2006} as only these currently provide spectra covering variations in [$\alpha$/Fe]. 

We compare our far-red index measurements with measurements from the SPS models of CvD12. These provide SSP spectra covering the wavelength range 1500--24000\AA\ with variations in the IMF slope (`bottom-light' through to $x=3$ `bottom-heavy'), age (3--13.5 Gyr), $\alpha$-element enhancement [$\alpha$/Fe] (0.0--+0.3), and variations in individual elemental abundance ratios [X/Fe] for 17 different elements (for fixed 13.5 Gyr age and fixed Chabrier IMF). The `base set' of CvD12 models (variations in age and IMF) are at solar metallicity, whereas the SSPs with variations in [$\alpha$/Fe] and [X/Fe] are at fixed [Fe/H]$=0.0$, so the total metallicity [Z/H] also changes according to the relation between metallicity, iron abundance and $\alpha$-enhancement from \citet{Trager2000b},
\begin{equation}
\label{eq:iron-alpha}
[{\rm Fe/H}] = [{\rm Z/H}] - A[\alpha/{\rm Fe}],
\end{equation}
where $A=0.93$ (Conroy, priv. comm.). However, these are not a complete representation of enhanced metallicity SSPs as only the $\alpha$-elements are increased. When we derive IMF slopes in Section~\ref{imf-sigma} we try to account for stellar population variations across the galaxies. We discuss the caveats of our approach and show the variations to be only second-order effects in Appendix~\ref{sec:imfcaveats}.

\subsection{Stellar populations from published optical measures}

The stellar populations of the three Coma BCGs have been well studied using optical spectra \citep[e.g. ][]{Jorgensen1999, Mehlert2000, Moore2002, Mehlert2003, Nelan2005, Sanchez-Blazquez2006c, Trager2008, Loubser2009, Coccato2010, Loubser2012, Groenewald2014}. \citet{Trager2008} and \citet{Loubser2009} summarise the ages, metallicities and $\alpha$-enhancements derived through Lick indices (H$\beta$, Mg{\it b}, Fe52, Fe53) for each galaxy from the literature. The values are derived using 2.7 arcsec equivalent apertures and thus represent the central regions of each galaxy.

\begin{figure*}
 \centering
 \includegraphics[width=12cm]{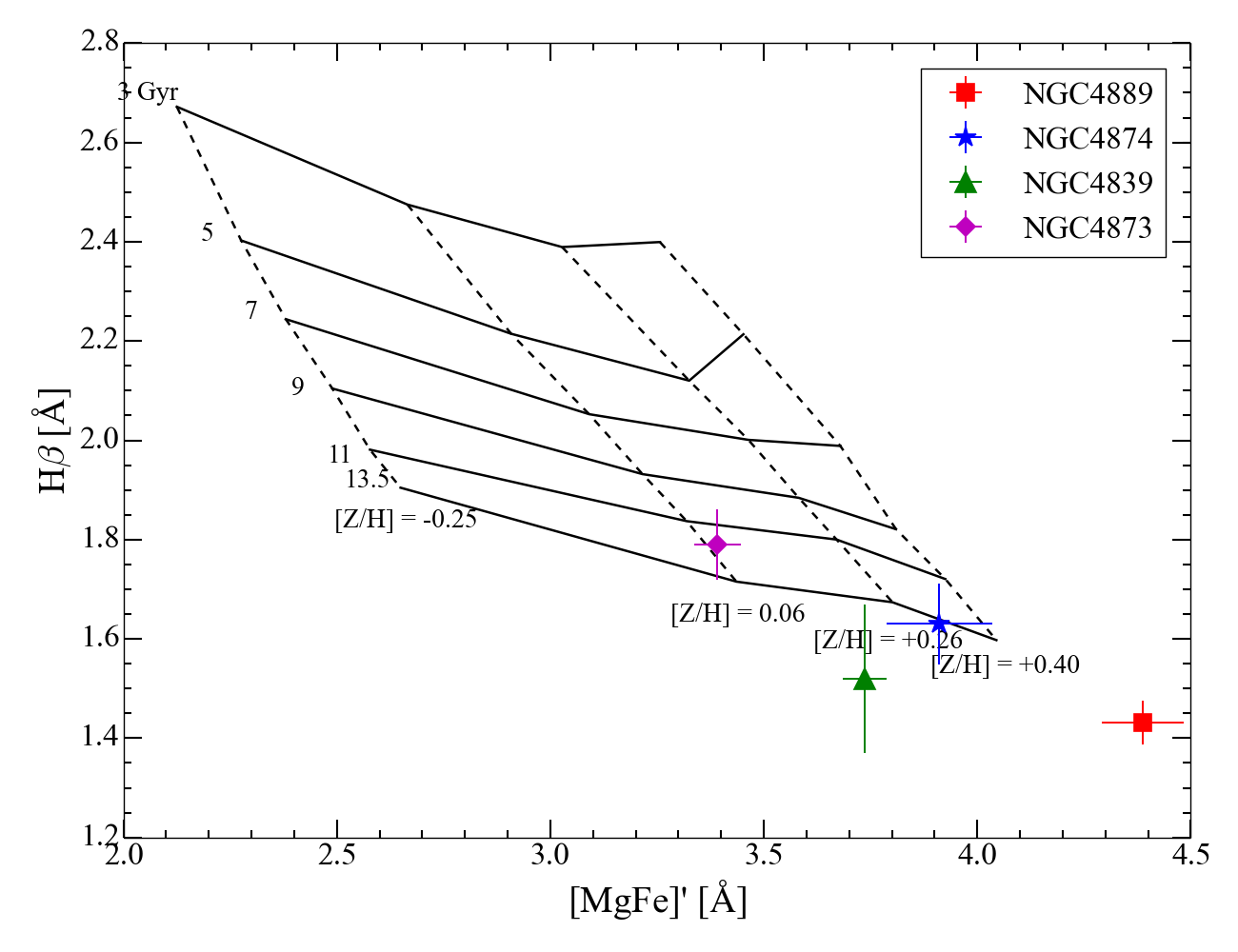}\\
  \includegraphics[width=12cm]{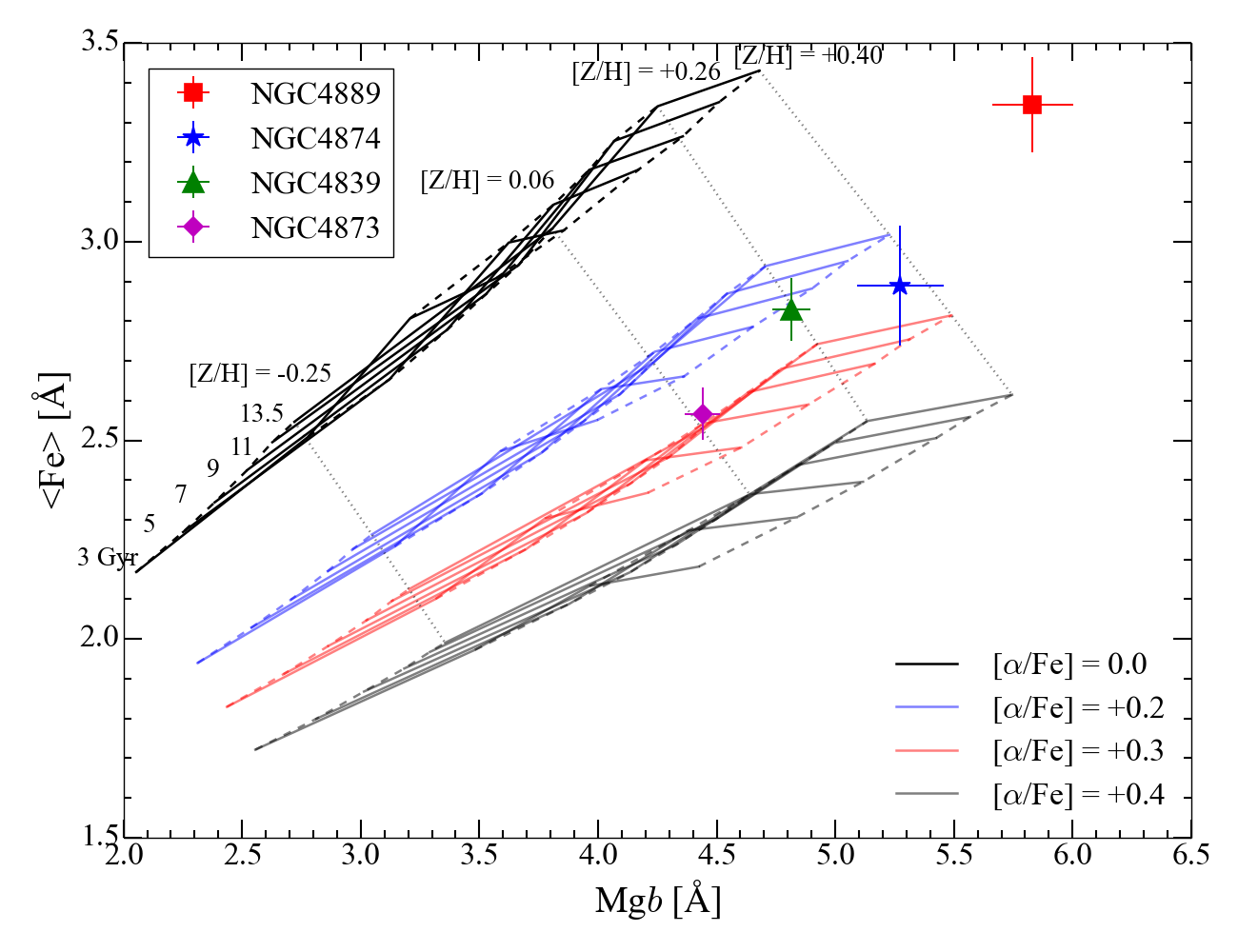}
 \caption{{\bf Top} shows the [MgFe]$'$--H$\beta$ index--index diagram showing positions of the four galaxies in our sample, compared with the SPS model predictions of \citet[][hereafter V15]{Vazdekis2015}. The data points are taken from published results in the literature; the data points for the three BCGs represent values from \citet{Trager2008} and \citet{Loubser2009}, measured from $2.7''$ aperture spectra and $R_{\rm e}/8$ aperture spectra respectively. The V15 models show variation in age and total metallicity for Kroupa IMF (both indices are insensitive to IMF). H$\beta$ traces variations in age and the [$\alpha$/Fe]-independent [MgFe]$'$ index traces total metallicity. The data points and models are both at a common velocity dispersion of $\sigma=200\,{\rm km}\,{\rm s}^{-1}$, corresponding to the Lick index resolution. All galaxies show similar ages of $\sim13.5\,{\rm Gyr}$, with NGC4889 showing the highest metallicity at [Z/H]$\,\sim0.5$, NGC4873 at [Z/H]$\,\sim0.3$, NGC4839 at [Z/H]$\,\sim0.25$ (note the weaker H$\beta$ strength possibly due to emission) and NGC4873 at [Z/H]$\,\sim0.06$. {\bf Bottom} shows the Mg{\it b}--<Fe> index--index diagram showing positions of the four galaxies in our sample, compared with the V15 SPS model predictions for a Kroupa IMF (both indices are again insensitive to IMF). The data points are taken from published results in the literature; the data points for the three BCGs represent values from \citet{Loubser2009}, and the point for NGC4873 represents the average from values in \citet{Trager2008}. The data points and models are both at a common velocity dispersion of $\sigma=200\,{\rm km}\,{\rm s}^{-1}$, corresponding to the Lick index resolution. All galaxies are $\alpha$-enhanced with [$\alpha$/Fe]$\sim+0.25$.}
 \label{fig:lickindices}
\end{figure*}

\citet{Trager2008} and \citet{Loubser2009} have presented Lick index measurements for the centres of the galaxies in our sample, derived from apertures of 2.7 arcsec and $R_{{\rm e}}/8$ in diameter respectively. In Fig.~\ref{fig:lickindices} we plot [MgFe]$'$--H$\beta$ (top) and Mg{\it b}--<Fe> (bottom) index--index diagrams and compare published optical measurements for the Coma galaxies to the V15 SPS models. <Fe> is a combination of the Fe52 and Fe53 indices, defined as,
\begin{equation}
{\rm <Fe>} =({\rm Fe52}+{\rm Fe53})/2,
\end{equation}
 and [MgFe]$'$ is a combination of the Fe indices with Mg{\it b}, defined by \citet{Thomas2005} as,
\begin{equation}
{\rm [MgFe]}' = \sqrt{({\rm Mg}{\it b})\times(0.72\times{\rm Fe}52 + 0.28\times{\rm Fe}53)}.
\end{equation}
All measurements are at a common resolution of $\sigma=200\,{\rm km}\,{\rm s}^{-1}$, corresponding to the Lick index resolution of these features \citep{Worthey1997}. The [MgFe]$'$--H$\beta$ diagram shows variations in age and metallicity from the V15 models as labelled on the subplot. These two indices are known to be reasonably independent of $\alpha$-enhancement and are completely independent of IMF. Comparing data to models, the galaxies are all consistently old at $\sim13.5\,{\rm Gyr}$. The metallicities range from [Z/H] $\sim +0.05$ for NGC4873, [Z/H] $\sim +0.25$ for NGC4839, [Z/H]$\sim+0.3$ for NGC4874, and [Z/H] $\sim+0.5$ for NGC4889.

The Mg{\it b}--<Fe> plot shows variations in age, metallicity and [$\alpha$/Fe] from the V15 models as labelled on the subplot. All the galaxies are consistent with an $\alpha$-enhancement of [$\alpha$/Fe] $\sim+0.25$. The metallicities are consistent with those derived from the [MgFe]$'$--H$\beta$ plot.

\subsection{Analysis of far red features}

\begin{figure*}
 \centering
 \includegraphics[width=17cm]{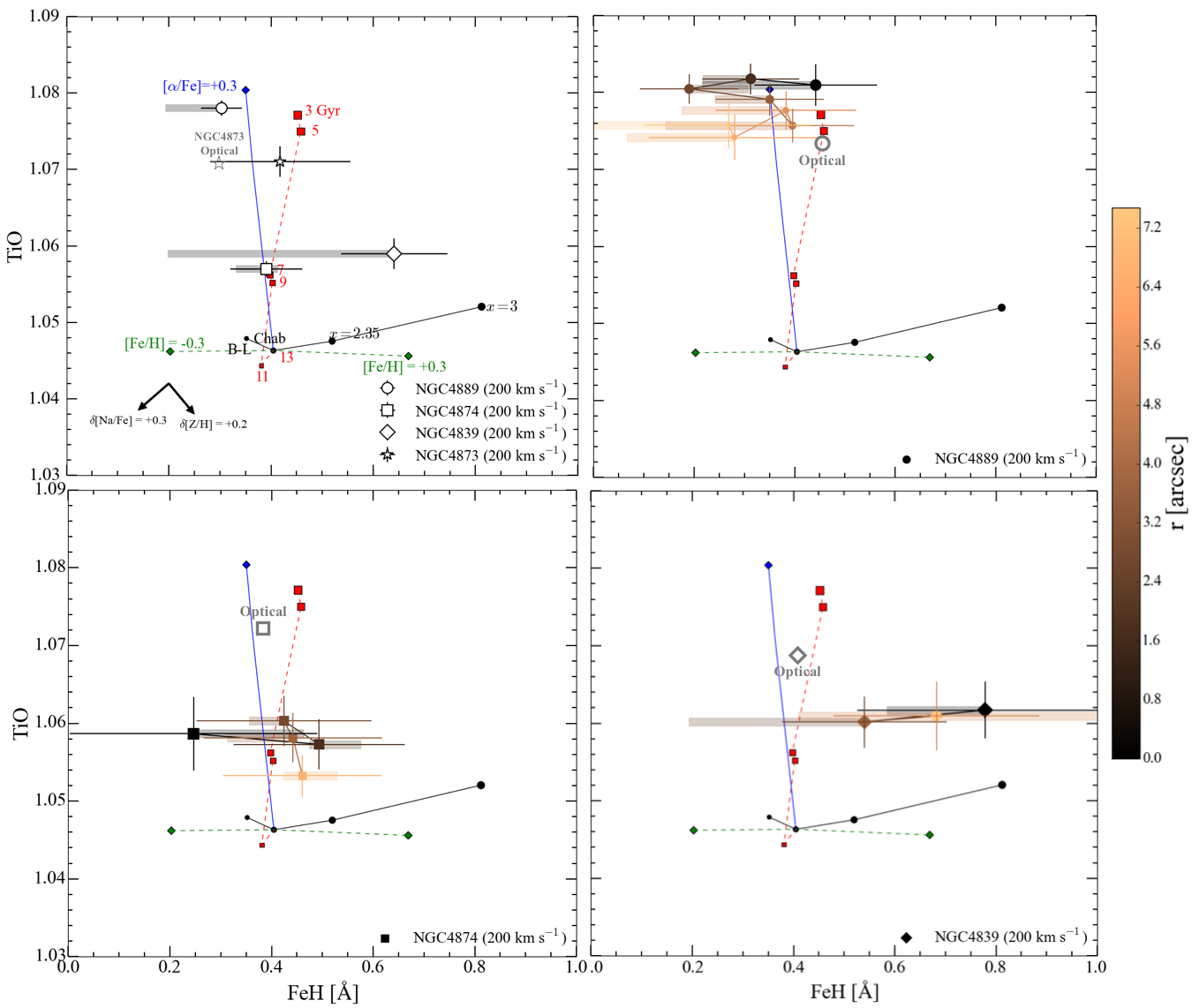}
 \caption{FeH--TiO index--index diagrams showing the global positions (top left sub-panel) and radial variation (other three sub-panels) of our galaxies as shown in the legends, compared with model predictions of CvD12 at 13.5 Gyrs. In the top left plot the white filled markers show the optimally extracted global spectrum measurements for each galaxy. Data and models are compared at a common resolution of $\sigma=200\,\rmn{km}\,\rmn{s}^{-1}$. In the bottom plot the colour bar and marker sizes indicate the radial position from the centre of each galaxy; size decreases and colour lightens with increasing radial distance. The FeH index measurements also show the systematic uncertainties as shaded rectangles. The CvD12 models show variation in IMF slope (black circles with each IMF labelled; B-L for bottom-light, and Chab for Chabrier), age (red squares), [$\alpha$/Fe] (blue diamonds), and individual elemental abundance variations for a 13.5 Gyr Chabrier IMF (green diamonds). The black arrows show the index responses to changes in: total metallicity as derived from the V15 models (the FeH response is approximated from the vector shown in Fig. 10 of \citet{LaBarbera2016}), and sodium abundance as derived from the CvD12 models. In each sub-panel, the grey marker labelled {\it Optical} shows the reference position for each galaxy based upon the optical indices from Fig.~\ref{fig:lickindices}, assuming a Chabrier IMF and with no over/under abundance of individual elements. For NGC4889 (circles), we infer an old, $\alpha$-enhanced population with a Chabrier, or possibly bottom-light, IMF. For NGC4874 we infer a Chabrier IMF, whereas for NGC4839 we see some evidence of a bottom-heavy ($x>2.35$) IMF in the centre, although with a very large systematic uncertainty. NGC4873 is also fully consistent with an Chabrier IMF.}
 \label{fig:CvD12_FeH_all_galaxies_200kms_corrfac}
\end{figure*}

\begin{figure*}
 \centering
 \includegraphics[width=17cm]{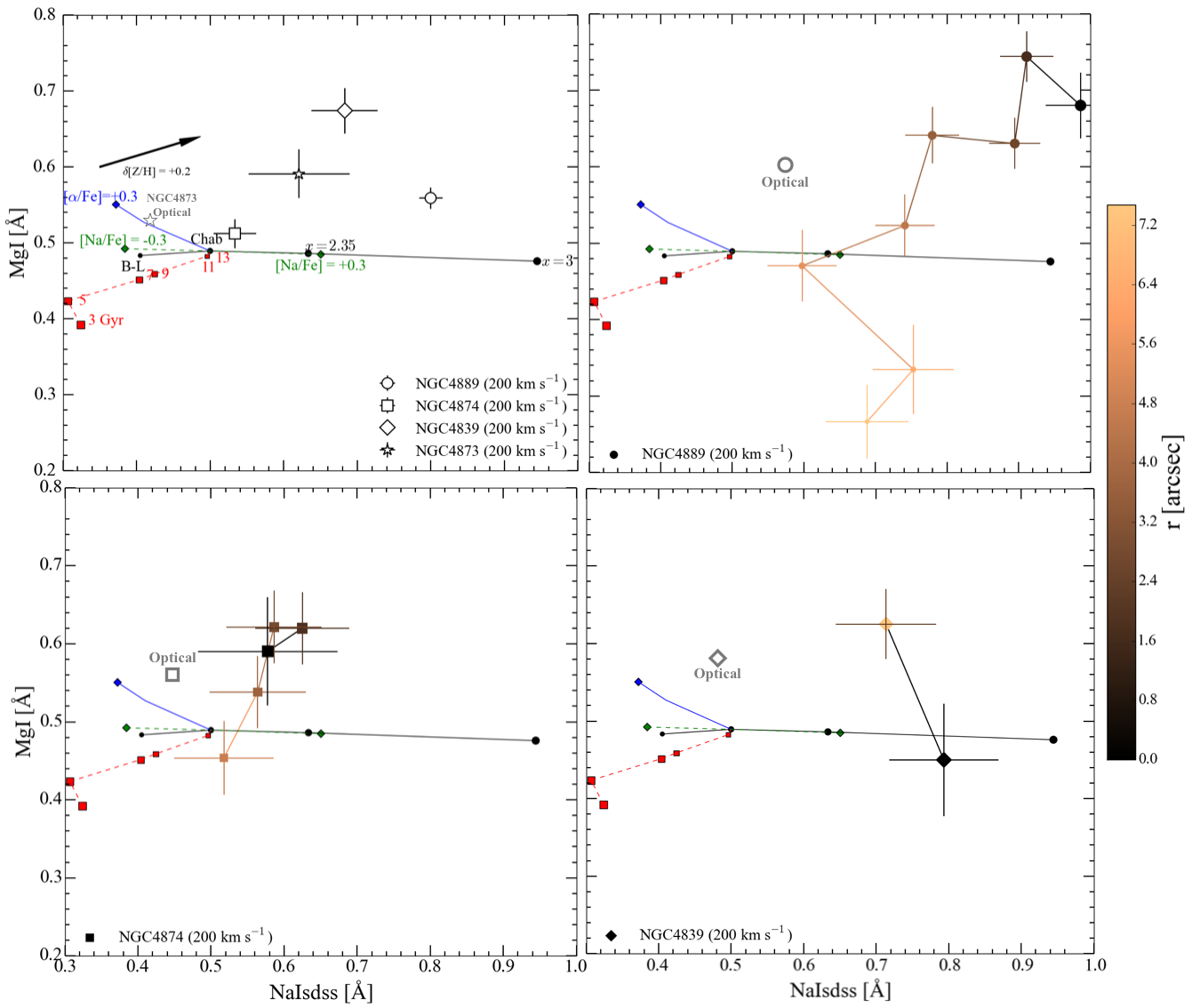}
 \caption{Same as Fig.~\ref{fig:CvD12_FeH_all_galaxies_200kms_corrfac} but for NaI$_{\rm{SDSS}}$--MgI indices: diagrams showing the global positions (top left sub-panel) and radial variation (other three sub-panels) of our galaxies as shown in the legends, compared with model predictions of CvD12 at 13.5 Gyrs. In each sub-panel, the grey marker labelled {\it Optical} shows the reference position for each galaxy based upon the optical indices from Fig.~\ref{fig:lickindices}, assuming a Chabrier IMF and with no over/under abundance of individual elements. For NGC4889 we fix the IMF as Chabrier from the FeH--TiO diagram and infer from the NaI$_{{\rm SDSS}}$--MgI diagram strong gradients in either the relative abundances of sodium and magnesium, or an overall metallicity gradient, rather than a bottom-heavy $x=3$ IMF. NGC4874 and NGC4873 are consistent with a Chabrier IMF and metal- and $\alpha$-enhanced populations. For NGC4839 the strong NaI$_{{\rm SDSS}}$ measurements provide some evidence of a bottom-heavy ($x>2.35$) IMF in the centre, although with the large FeH uncertainty this is also consistent with Na-enhancement.}
 \label{fig:CvD12_NaI_all_galaxies_200kms_corrfac}
\end{figure*}

In Figs.~\ref{fig:CvD12_FeH_all_galaxies_200kms_corrfac} and \ref{fig:CvD12_NaI_all_galaxies_200kms_corrfac} we plot FeH--TiO and NaI$_{\rm{SDSS}}$--MgI index-index diagrams showing 13.5 Gyr CvD12 model predictions (labelled) and our galaxy measurements (labelled in legend) at a common resolution of $200\,\rmn{km}\,\rmn{s}^{-1}$. From the FeH--TiO diagram we see from the CvD12 models that these indices work in orthogonal directions; FeH positively correlates with IMF slope and [Fe/H], whereas TiO increases with increasing [$\alpha$/Fe]. The NaI$_{\rm{SDSS}}$--MgI diagram shows that MgI is independent of IMF, positively correlates with [Z/H] and negatively correlates with [$\alpha$/Fe]. NaI$_{\rm{SDSS}}$ is sensitive to IMF, $\alpha$- and sodium abundance. We now discuss the analysis of each galaxy in turn.

\subsection{NGC4889}

For NGC4889, from Fig.~\ref{fig:lickindices} and Equation~\ref{eq:iron-alpha} we adopt an age of 13.5 Gyr, metallicity of [Z/H] = +0.5, $\alpha$-enhancement of [$\alpha$/Fe] = +0.25, and iron-enhancement of [Fe/H] = +0.27.

Fig.~\ref{fig:CvD12_FeH_all_galaxies_200kms_corrfac} shows the NGC4889 data points with circle markers. In the FeH-TiO subplot the NGC4889 measurements scatter around the [$\alpha$/Fe] = +0.3, 13.5 Gyr Chabrier IMF SSP, with no radial trends. The FeH measurements reject a bottom-heavy $x=3$ IMF slope and the global spectrum measurement strongly rejects a bottom-heavy IMF. In combination with the optical indices, our measurements qualitatively suggest that NGC4889 is consistent with an old, $\alpha$-enhanced population with a Chabrier, or possibly even bottom-light, IMF.

The NaI$_{\rm{SDSS}}$--MgI diagram in Fig.~\ref{fig:CvD12_NaI_all_galaxies_200kms_corrfac} shows the radial gradients of these two indices in NGC4889. As we take the age, [$\alpha$/Fe] and IMF to be fixed from the optical indices and FeH--TiO diagram, we infer a strong negative gradient in the sodium abundance, or a total metallicity gradient \citep[e.g.][]{Coccato2010}, rather than the interpretation of a bottom-heavy $x=3$ IMF with decreasing age gradient.

The most striking behaviour from the radial index measurements is the discrepancy between the large NaI$_\rmn{SDSS}$ gradient and the flat FeH profile. \citet{Zieleniewski2015} and \citet{McConnell2016} have both measured the same discrepant behaviour, in M31 and two massive ETGs respectively. These galaxies have lower central velocity dispersions between $200<\sigma_*<250\,\rmn{km}\,\rmn{s}^{-1}$ and so NGC4889 represents a much higher dispersion and more massive galaxy where bottom-heavy IMFs have been recently proposed \citep[e.g.][]{Cappellari2012, ConroyVanDokkum2012b, LaBarbera2013, Ferreras2013, Spiniello2014}. Furthermore, \citet{VanDokkumConroy2010} presented a stacked spectrum around NaI from four galaxies in the Coma cluster, which included NGC4889, and they concluded a bottom-heavy IMF. However, they did not present the stacked Coma spectrum around FeH. Thus, the flat FeH profile we measure presents tension with the notion of increased IMF slope in this galaxy and with the general IMF-$\sigma_*$ relation (see Section~\ref{imf-sigma}).

A Chabrier IMF gives a sodium enhancement in the centre of [Na/Fe]$\sim+1.0$, decreasing to ${\rm [Na/Fe]}\,\sim+0.3$ at 5 arcsec (2.5 kpc). An overall metallicity trend would give ${\rm [Z/H]}\,\sim+0.6$ in the centre down to ${\rm [Z/H]}\,\sim+0.3$ at 10 arcsec, fully consistent with \citet{Coccato2010}. We note from the [Na/Fe] vector on Fig.~\ref{fig:CvD12_FeH_all_galaxies_200kms_corrfac} that Na-enhancement slightly suppresses the measured FeH strength, so the strong central Na-enhancement would act to weaken the central FeH measurement in NGC4889. However, this is more than offset by the [Fe/H] vector and we have accounted for both effects when measuring the IMF slopes in Section~\ref{imf-sigma} (see Appendix~\ref{sec:imfcaveats} for further details).

\subsection{NGC4874}

For NGC4874, from Fig.~\ref{fig:lickindices} and Equation~\ref{eq:iron-alpha} we adopt an age of 13.5 Gyr, metallicity of [Z/H] = +0.25, $\alpha$-enhancement of [$\alpha$/Fe] = +0.25, and iron-enhancement of [Fe/H] = +0.02.

Figs.~\ref{fig:CvD12_FeH_all_galaxies_200kms_corrfac} and \ref{fig:CvD12_NaI_all_galaxies_200kms_corrfac} show the NGC4874 data points with square markers. We have excluded the anomalously high FeH datapoint at $r=5\,\rm{arcsec}$ and the outermost NaI$_{{\rm SDSS}}$ datapoint from these figures for clarity. The remaining FeH--TiO points scatter around the Chabrier IMF slope with a bottom-heavy $x=3$ IMF excluded. The TiO index places the galaxy at around [$\alpha$/Fe]$\sim+0.1$ for an 13.5 Gyr population. Similarly to NGC4889, we find from the FeH index that NGC4874 is consistent with a Chabrier IMF and solar iron abundance, but we cannot rule out the possibility of a dwarf-dominated IMF and iron deficiency. The NaI$_{\rm{SDSS}}$--MgI diagram again suggests either a weak gradient in magnesium or an overall metallicity gradient. The flat profile of NaI$_{\rm{SDSS}}$ at $\sim0.4$\AA\ (excluding the outermost point which has a bad red continuum region) suggests a sodium enhancement of ${\rm [Na/Fe]}\sim+0.2$ for a Chabrier IMF. The NaI$_{\rm{SDSS}}$ and FeH strengths at $\sim0.4$\AA\ strengthen the case for a Chabrier IMF in this galaxy, rather than a bottom-heavy IMF, which would require sodium deficiency to match the observed NaI$_{\rm{SDSS}}$ profile.

\subsection{NGC4839}

For NGC4839, from Fig.~\ref{fig:lickindices} and Equation~\ref{eq:iron-alpha} we adopt an age of 13.5 Gyr, metallicity of [Z/H] = +0.3, $\alpha$-enhancement of [$\alpha$/Fe] = +0.25, and iron-enhancement of [Fe/H] = 0.07.

Figs.~\ref{fig:CvD12_FeH_all_galaxies_200kms_corrfac} and \ref{fig:CvD12_NaI_all_galaxies_200kms_corrfac} show the NGC4839 data points with diamond markers. The FeH measurements scatter around a bottom-heavy $x>2.35$ IMF slope and the TiO measurements place the galaxy around [$\alpha$/Fe] $\sim+0.1$ for a 13 Gyr population. The TiO measurements are very similar to NGC4874. The key difference for this galaxy is the potentially deeper FeH absorption compared with the other two BCGs, although we note that there are large systematic uncertainties on the FeH measurements, and the central measurement shows a deeper strength at only the 1$\sigma$ level. The global spectrum FeH measurement places the galaxy at an IMF slope of $x\sim2.5$ at the 1$\sigma$ level. From the NaI$_{\rm{SDSS}}$--MgI diagram we infer that NGC4839 is also metal enhanced, and the strong NaI$_{\rm{SDSS}}$ index with no gradient also supports a possible steeper IMF slope, but is equally consistent with an Na-enhancement of ${\rm [Na/Fe]}\,\sim+0.5$ for a Chabrier IMF.

\subsection{NGC4873}

For NGC4873, from Fig.~\ref{fig:lickindices} and Equation~\ref{eq:iron-alpha} we adopt an age of 12 Gyr, metallicity of [Z/H] = +0.06, $\alpha$-enhancement of [$\alpha$/Fe] = +0.25, and iron-enhancement of [Fe/H] = -0.17.

Figs.~\ref{fig:CvD12_FeH_all_galaxies_200kms_corrfac} and \ref{fig:CvD12_NaI_all_galaxies_200kms_corrfac} show the NGC4873 global data point with the star marker. Our measurements are consistent with NGC4873 being old and $\alpha$-enhanced. The FeH measurement is consistent with a Chabrier IMF from FeH. This is further supported by the NaI$_{\rm{SDSS}}$ measurement if a sodium enhancement of ${\rm [Na/Fe]}\sim+0.4$ is invoked.

\section{Discussion}

\subsection{The IMF-$\sigma_*$ relation}
\label{imf-sigma}

\citet{Ferreras2013}, \citet{LaBarbera2013} and \citet{Spiniello2014} have all published IMF-$\sigma_*$ relations for ETGs showing that an increase in $\sigma_*$ corresponds to an increase in IMF slope. These relations are all derived using galaxies covering a range of velocity dispersions from $\sim150$--$320\,\rmn{km}\,\rmn{s}^{-1}$, and so all fall well within this range. While there is some variation between the relations, the general conclusion is that galaxies with $\sigma_*>200\,\rmn{km}\,\rmn{s}^{-1}$ (for LB13 and F13; $\sigma_*>250\,\rmn{km}\,\rmn{s}^{-1}$ for S14) have IMFs steeper than Salpeter, and galaxies with $\sigma_*<200\,\rmn{km}\,\rmn{s}^{-1}$ have IMFs approaching Chabrier. Therefore our qualitative analysis for NGC4873 ($\sigma_*\sim200\,\rmn{km}\,\rmn{s}^{-1}$) having a normal IMF agrees well with these relations. NGC4874 and NGC4839 have $\sigma_*\sim270\,\rmn{km}\,\rmn{s}^{-1}$ and the index interpretation of NGC4839 having a steeper (than Salpeter) IMF slope also agrees. However, our global measurements for NGC4874 are not consistent with an IMF slope steeper than Salpeter. Finally, NGC4889 ($\sigma_*\sim400\,\rmn{km}\,\rmn{s}^{-1}$) has a central dispersion much larger than the galaxies used to derive the IMF-$\sigma_*$ relations, but our inferred Chabrier IMF from the flat FeH profile is inconsistent with the relations, which suggest a very bottom-heavy IMF slope ($x\sim3$).

\begin{figure*}
\centering
\includegraphics[width=16cm]{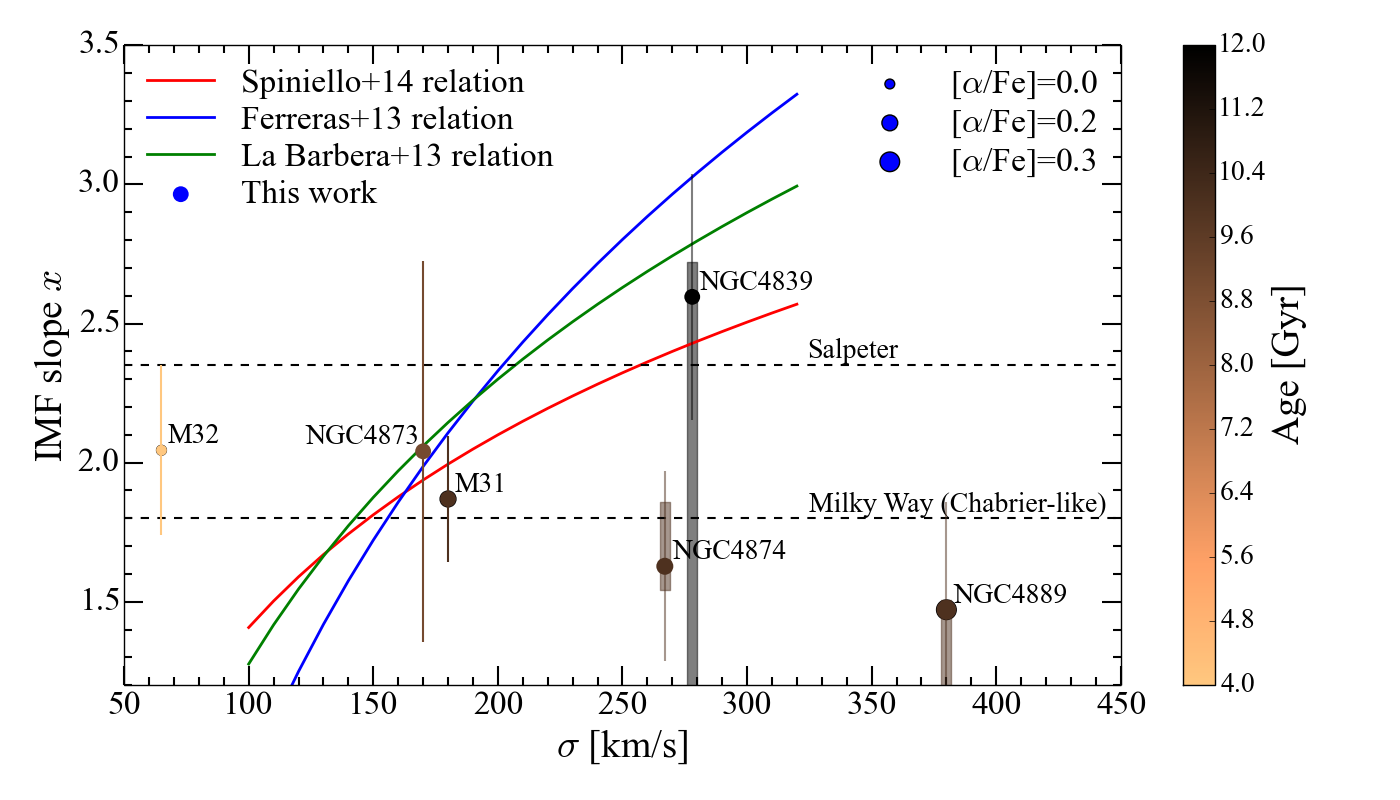}
\caption{Derived IMF slope $x$ against central velocity dispersion for our sample of Coma galaxies, as well as M32 and M31 from \citet{Zieleniewski2015}. The IMF slope has been derived from our {\it global} FeH measurements accounting for CvD12 model responses to age, [$\alpha$/Fe] and [Fe/H] variations, as described in Section~\ref{imf-sigma}, using published values in the literature for each galaxy \citep{Rose2005, Trager2008, Loubser2009, Saglia2010, Coccato2010, Loubser2012}. For our data, the colour of each point indicates the age and the marker size indicates the level of [$\alpha$/Fe] enhancement for each galaxy. The coloured lines show {\bf unimodal} IMF-$\sigma_*$ relations from the previous works of \citet{Ferreras2013, LaBarbera2013} and \citet{Spiniello2014}, derived using central $2.7''$ SDSS spectra and several optical/NIR features, including NaD/NaI but excluding FeH. The dashed horizontal lines show the Salpeter and Milky-Way (Chabrier-like) slopes. Our results show consistency with a universal Chabrier IMF across a large range of velocity dispersions, with the exception of the single galaxy NGC4839, for which we see possible evidence of an IMF slightly heavier than Salpeter, albeit with a very large systematic uncertainty.}
\label{all_galaxies_imf_constraints}
\end{figure*}

To quantitatively compare our results to the published IMF-$\sigma_*$ relations we derive single power-law IMF slopes for our full sample of galaxies using the CvD12 models. When deriving the IMF slopes we take into account the derived values for age, [Z/H], [$\alpha$/Fe], [Fe/H] and [Na/Fe]. The method is described in Appendix~\ref{sec:imfcaveats}. Our derived IMF slopes for each galaxy are shown in Fig.~\ref{all_galaxies_imf_constraints} as a function of velocity dispersion. We find consistency with a Chabrier IMF for each galaxy across the $\sigma_*$ range, except for NGC4839 where we find possible evidence of an IMF steeper than Salpeter, although with a very large uncertainty. We also show the IMF-$\sigma_*$ relations derived by \citet{Ferreras2013}, \citet{LaBarbera2013} and \citet{Spiniello2014} using optical Na and TiO features. We discuss the caveats of our IMF derivations in Appendix~\ref{sec:imfcaveats}. However we note that there is negligible change in our derived values for $x$ whether we account for stellar population parameter variations between galaxies, or not.

\citet{Ferreras2013} and \citet{LaBarbera2013} have also invoked a bimodal form of the IMF and fitted it to observational data to derive IMF$_{\rm{b}}-\sigma_*$ relations. The bimodal IMF is defined in \citet{Vazdekis1996} with a slope of $x$ for $M>0.6\,\rmn{M}_{\odot}$ and flattening below $0.6\,\rmn{M}_{\odot}$. For an exponent of $x=2.3$ it very closely traces the \citet{Kroupa2001} universal IMF. \citet{LaBarbera2016} recently used the FeH index to distinguish between bottom-heavy unimodal and bimodal IMFs in a massive ETG ($\sigma_*\sim300\,\rmn{km}\,\rmn{s}^{-1}$). FeH is most sensitive to $M\leq0.3M_\odot$ and thus can constrain the very low mass end of the IMF. From \citet{LaBarbera2016} the FeH index is predicted to vary much more weakly with increasing $x$, compared to the unimodal IMF. In fact a bottom-heavy bimodal IMF with $x=4$ predicts a similar FeH strength to the Salpeter IMF \citep[at $\sigma_*=300\,\rmn{km}\,\rmn{s}^{-1}$; see Figure 10 in][]{LaBarbera2016} of $\sim0.45$\AA. However, NGC4889 has an even weaker FeH feature at a larger $\sigma_*$ and seems to be incompatible with the specific, extreme bimodal bottom-heavy IMF slope of $x=4$ (for $M>0.6\,{\rm M}_\odot$) proposed in \citet{LaBarbera2016}.

\subsection{Comparison with dynamical modelling and lensing results}

Stellar $M_*/L_{R}$ have been published for each of the BCGs by \citet{Thomas2007, Thomas2011} and for NGC4889 by \citet{McConnell2012}. For NGC4889, the result by \citet{Thomas2007} of $M_*/L_{R}=6.5\,\rm{M_{\odot}}\,\rm{L_{\odot}}^{-1}$ (\citeauthor{Thomas2007} do not quote 1$\sigma$ uncertainties) and by \citet{McConnell2012} of $M_*/L_{R}=5.9\pm1.7\,\rm{M_{\odot}}\,\rm{L_{\odot}}^{-1}$ are in good agreement. For NGC4874 and NGC4839, \citet{Thomas2007, Thomas2011} measured $M_*/L_{R}=7.0\,\rm{M_{\odot}}\,\rm{L_{\odot}}^{-1}$ and $M_*/L_{R}=8.5\pm2.0\,\rm{M_{\odot}}\,\rm{L_{\odot}}^{-1}$ respectively. It should be noted that these represent upper limits on $M_*/L$ due to the uncertainties of incorporating a dark matter component in the models \citep[e.g.][]{Thomas2011, Bundy2015}.

The order of increasing $M_*/L_{R}$ between NGC4889, NGC4874 and NGC4839 is echoed by the IMFs we infer from the FeH measurements within these galaxies. Calculating $M/L_{R}$ from the best-fit SSPs used to derive the IMF slope for each galaxy, for NGC4889 we measure $M_*/L_{R}=7.2\pm2.0\,\rm{M_{\odot}}\,\rm{L_{\odot}}^{-1}$, for NGC4874 we measure $M_*/L_{R}=5.5\pm1.0\,\rm{M_{\odot}}\,\rm{L_{\odot}}^{-1}$, and for NGC4839 we measure $M_*/L_{R}=8.7\pm1.5\,\rm{M_{\odot}}\,\rm{L_{\odot}}^{-1}$. These mass-to-light ratios from SSP models are in general agreement with the published dynamical measurements.

Mass-to-light ratios measured from gravitationally lensed galaxies also provide an interesting comparison. \citet{SmithLucey2013} have presented a massive lensed ETG ($\sigma_*\approx330\,\rm{km}\,\rm{s}^{-1}$) with a mass excess factor $\alpha=(M/L)_{\rm{lens}}/(M/L)_{\rm{kroup}}=1.04\pm0.15$, which strongly favours a Kroupa or Chabrier IMF and rejects even a Salpeter IMF. Furthermore, \citet{Smith2015} have presented an average mass excess factor derived from three gravitationally lensed massive ETGs ($\sigma_*>300\,\rm{km}\,\rm{s}^{-1}$). They find $\langle\alpha\rangle=1.1$, which again rejects a Salpeter IMF and strongly rejects bottom-heavy IMFs in these galaxies. Thus, our results are consistent with IMF determinations for other massive ETGs using independent methods.

\subsection{IMF or abundance gradients and implications for the stellar populations of BCGs}

Strong radial gradients of absorption features have now been documented at small radii in some ETGs by several authors, most notably for sodium \citep[e.g.][]{FaberFrench1980, Boroson1991, Martin-Navarro2015a, Martin-Navarro2015c, Zieleniewski2015, McConnell2016}. Furthermore, the discrepancy between the strengths of sodium and iron `IMF' indices has also been noted by numerous authors \citep[e.g.][]{Smith2012, ConroyVanDokkum2012a, SmithLucey2013, Zieleniewski2015, McConnell2016}. This poses interesting questions for understanding the physical context of these gradients. The general scenario is that BCGs form in the deepest potential wells in short, intense star formation bursts, with successive growth through small-scale mergers around the outskirts \citep[e.g.][]{Larson1974, Carlberg1984, Naab2009, Hopkins2010, Cappellari2016}. This inside-out formation scenario predicts metallicity gradients, as the central stellar populations enrich in-falling gas sustaining the ongoing star formation. For NGC4889 we infer a total metallicity, or magnesium and sodium abundance, gradients. Tracing the radial behaviour of the NaD feature would help further constrain the underlying physical driver behind the NaI$_{{\rm SDSS}}$ gradient. However, sodium is not an $\alpha$-element and is generally poorly understood in the context of galaxy stellar populations \citep[e.g.][]{Jeong2013, Spiniello2014, Smith2015b}, so further study of sodium-enhancement in galaxies is required.

The recent study of IMF variations has shown correlations with both [$\alpha$/Fe] and central velocity dispersion \citep{Cappellari2012, ConroyVanDokkum2012b, Ferreras2013, LaBarbera2013, Spiniello2014}. One physical model is that larger mass and $\alpha$-enhanced systems have more type II supernovae driven turbulence, leading to greater fragmentation of star forming gas clouds \citep{ConroyVanDokkum2012b}. \citet{Chabrier2014} highlights that the IMF slope should correlate with {\it density} rather than velocity dispersion. Therefore, galaxies that underwent more rapid formation periods via mergers or intense gas inflows should have more bottom-heavy IMFs. NGC4839 sits in a sub-cluster located to the south west of the main Coma cluster. Our FeH measurements for this galaxy hint at the possibility of a different formation density compared to the other two BCGs, and so this would make an exciting object for follow-up observations targeting different IMF-sensitive features.

The very low FeH measured in NGC4889 could be consistent with a bottom-light IMF dominated by remnants rather than low-mass stars. This form has been proposed by studies of sub-millimetre galaxies with extremely high star formation rates, which have been thought to be the progenitors of BCGs \citep[e.g. ][c.f. \citealt{Hayward2013}]{Baugh2005, Narayanan2012, Narayanan2013}. Indeed, the dynamical modelling of \citet{Cappellari2012, Cappellari2013} shows a systematic increase of the $M_*/L$ with velocity dispersion, but cannot distinguish between bottom-light or bottom-heavy forms, so this scenario would still be consistent.

Our work, and other studies using independent methods \cite[e.g. gravitational lensing,][]{SmithLucey2013, Smith2015}, points to the notion that careful mass determinations of individual galaxies should be taken on a galaxy-by-galaxy basis. High S/N optical spectroscopy of these objects covering other known IMF-sensitive features \citep[see e.g.][]{Spiniello2014} would help elucidate the stellar component of these galaxies. Our results show that the Coma BCGs present an exciting target for IFS surveys like MaNGA \citep{Bundy2015}, which would be able to provide mass constraints using both, visible to far red spectral coverage, and dynamical modelling techniques.

\section{Conclusions}

Using the Oxford SWIFT instrument we have undertaken a study of the three brightest cluster galaxies in the Coma core, and south-west extension, with the aim of measuring radial gradients in IMF-sensitive far red stellar absorption features. We obtained high S/N integral field spectroscopy of NGC4889, NGC4874 and NGC4839, covering the central $\sim10\,\rmn{arcsec}\,(5\,\rmn{kpc})$ of each galaxy, as well as unresolved data for the fast rotator NGC4873. We developed two separate sky subtraction methods to minimise sky line residuals in our spectra and validate our results. We investigated the far red NaI doublet, calcium triplet CaT, magnesium MgI, titanium oxide TiO and iron hydride Wing-Ford band FeH as functions of radius and conclude:

1. NGC4889 shows strong negative gradients in NaI$_{\rmn{SDSS}}$ and CaT, mild negative gradients in MgI and TiO, and a flat profile of weak FeH absorption. The flat FeH profile contrasts with the strong NaI$_{\rmn{SDSS}}$ gradient in the context of spatial IMF variations. Comparing FeH and TiO indices to stellar population models we infer a Chabrier or possibly bottom-light IMF and old, $\alpha$-enhanced population, in agreement with optical stellar population studies. This result is especially interesting due to the large velocity dispersion of $\sim400\,\rmn{km}\,\rmn{s}^{-1}$ in the centre of NGC4889, which from previous IMF-$\sigma_*$ relations would infer a very bottom-heavy IMF.

2. NGC4874 shows weak negative gradients in CaT, MgI and TiO. NaI$_{{\rm SDSS}}$ and FeH profiles are both flat and do not show strong absorption. Comparing to models we again find a Chabrier IMF with lower $\alpha$-enhancement compared to NGC4889 and in agreement with optical stellar population studies.

3. NGC4873 displays strong CaT, MgI and TiO absorption. Our measurements of FeH and NaI$_{\rmn{SDSS}}$ suggest a Chabrier IMF, which is what we expect for a galaxy of this velocity dispersion.

4. Within NGC4839 we measure strong NaI {\it and} strong FeH absorption, although we note a large systematic uncertainty on FeH due to the sky subtraction process. There is possible evidence of a more dwarf-heavy population in the centre, consistent with Salpeter, but the large uncertainty prevents a strong conclusion. We compare to dynamical modelling results, which suggest an IMF up to $\sim1.5$--2.5 times heavier than Kroupa is consistent, which would be in agreement with a Salpeter IMF.

5. These galaxies span a large range of velocity dispersions and we find tension with the notion of a simple IMF-$\sigma_*$ relation compared to the interpretation from the IMF-sensitive features. Specifically, NGC4889 has a very high central dispersion ($\sim400\,\rmn{km}\,\rmn{s}^{-1}$) and we only detect low levels of FeH compared with a strong NaI gradient. We conclude that interpretations of dwarf-dominated stellar populations should be treated on a galaxy-by-galaxy basis. Combining near-infrared and optical IMF-sensitive spectral indices with dynamical models should give strong constraints on the dwarf-abundance in individual galaxies.

\section*{Acknowledgments}

We are grateful to the anonymous referee for providing feedback that greatly improved this manuscript. The Oxford SWIFT integral field spectrograph was supported by a Marie Curie Excellence Grant from the European Commission (MEXT-CT-2003-002792, Team Leader: N. Thatte). It was also supported by additional funds from the University of Oxford Physics Department and the John Fell OUP Research Fund. Additional funds to host and support SWIFT at the 200-inch Hale Telescope on Palomar were provided by Caltech Optical Observatories.

This work was supported by the Astrophysics at Oxford grants (ST/H002456/1 and ST/K00106X/1) as well as visitors grant (ST/H504862/1) from the UK Science and Technology Facilities Council. SZ is supported by STFC-HARMONI grant ST/J002216/1. RCWH was supported by the Science and Technology Facilities Council [STFC grant numbers ST/H002456/1, ST/K00106X/1 \& ST/J002216/1]. RLD acknowledges travel and computer grants from Christ Church, Oxford, and support from the Oxford Centre for Astrophysical Surveys, which is funded through generous support from the Hintze Family Charitable Foundation.

\bibliography{../../MASTERBIB}
\bibliographystyle{mnras}

\appendix

\section{Correcting to a common velocity dispersion}
\label{sec:corrsig}

\begin{figure*}
 \centering
 \includegraphics[width=11cm]{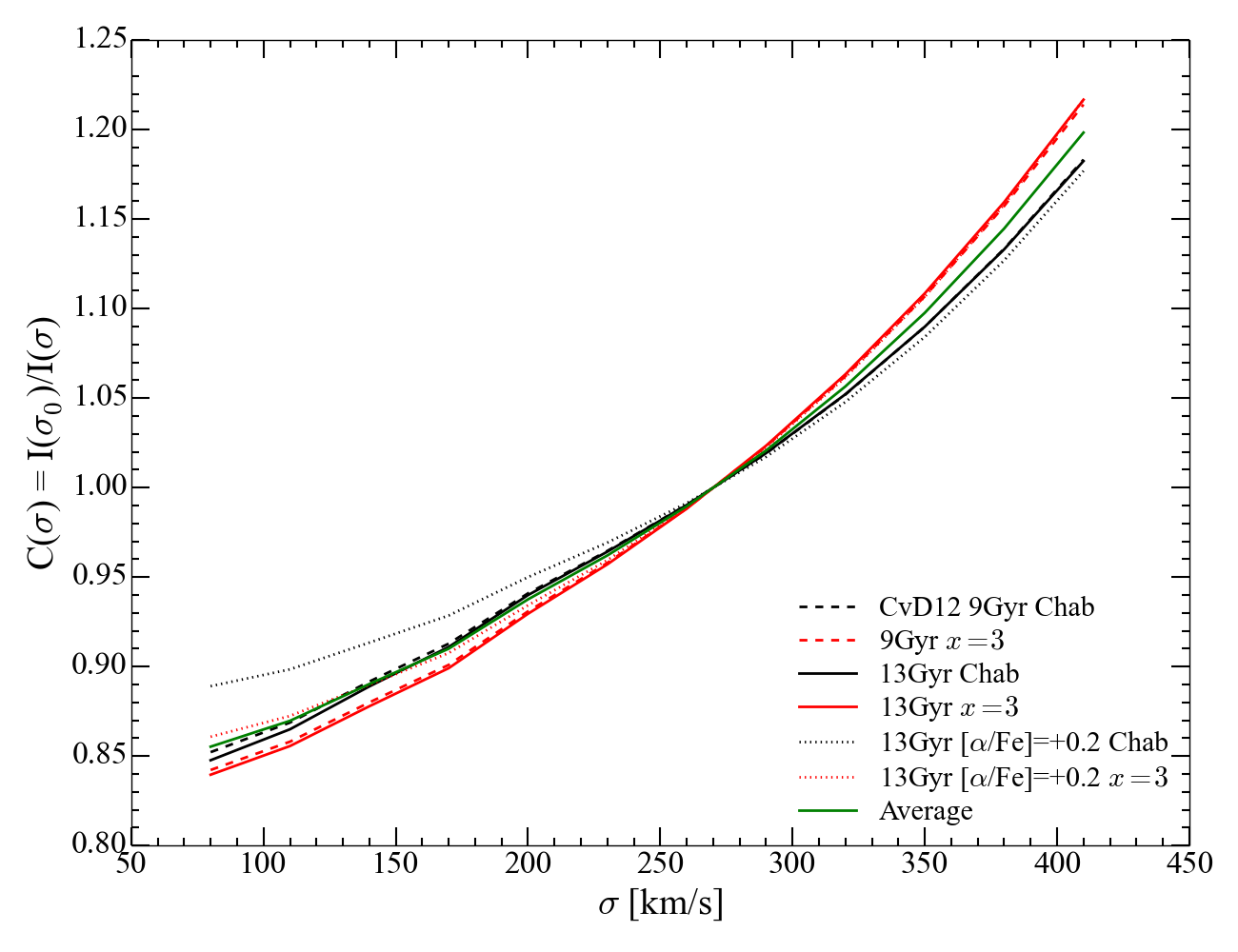}
 \caption{Plot showing the correction factor $C(\sigma)$ as a function of velocity dispersion $\sigma$ for the FeH index from a subset of CvD12 SSP spectra as shown in the legend. The correction factor is normalised to be unity at the intrinsic velocity dispersion of the galaxy and then the intrinsic galaxy index measurement is divided by the average value at the `correcting' dispersion, for which in this paper we use $\sigma=200\,{\rm km}\,{\rm s}^{-1}$. This plot shows the curves normalised at a value of $\sigma=270\,{\rm km}\,{\rm s}^{-1}$, which then gives an value of $C(\sigma_{200})=0.93$.}
 \label{fig:correctionfactor}
\end{figure*}

In order to equally compare our equivalent width measurements within each galaxy, as well as between galaxies, we apply the following procedure to correct for varying velocity dispersions. We first make index measurements $I(\sigma_*)$ at the intrinsic velocity dispersions of each spectrum. For each feature we use the CvD12 models to derive a correction factor, which is applied to the index measurement to scale it to a specific velocity dispersion. For each model SSP spectrum, we convolve it up to a range of $\sigma$ from 80--$410\,{\rm km}\,{\rm s}^{-1}$ and measure the index values at each $\sigma$. These $I(\sigma)$ curves are then normalised to be unity at the velocity dispersion of the galaxy spectrum $\sigma_*$, so we then take the average value of the correction factor,
\begin{equation}
C(\sigma)=I(\sigma_*)/I(\sigma),
\end{equation}
at the velocity dispersion we are correcting to and divide each galaxy index measurement by its corresponding factor. We use the CvD12 SPS models to derive the correction factors $C(\sigma)$ for our index measurements. Fig.~\ref{fig:correctionfactor} shows $C(\sigma)$ as a function of velocity dispersion $\sigma$ for the FeH index from a subset of CvD12 SSP spectra as shown in the legend. The correction factor is normalised to be unity at the intrinsic velocity dispersion of the galaxy spectrum and then the galaxy intrinsic index measurement is divided by the average value at the `correcting' dispersion, for which in this paper we use $\sigma=200\,{\rm km}\,{\rm s}^{-1}$. Fig.~\ref{fig:correctionfactor} shows the curves normalised at a value of $\sigma=270\,{\rm km}\,{\rm s}^{-1}$, which then gives an value of $C(\sigma_{200})=0.93$.

The CvD12 SSPs cover a range of stellar population parameters that appropriate for the galaxies studied in this paper. These include Chabrier and $x=3$ bottom-heavy IMFs, 9 and 13.5 Gyr ages, and solar and $\alpha$-enhanced (and therefore metal-enhanced) abundances. The spread of the curves introduces an additional uncertainty when correcting the index measurements to a different $\sigma$. Table~\ref{tab:corrsigerrors} shows the maximum additional uncertainty for each galaxy when correcting its central spectrum's indices, which have the largest velocity dispersions within each galaxy, to $\sigma=200\,{\rm km}\,{\rm s}^{-1}$.

\begin{table*}
\centering
  \caption{Maximum additional error and ($^{{\rm max}}_{{\rm min}}$) of $C(\sigma_{200})$ from correcting to a common velocity dispersion of $\sigma=200\,{\rm km}\,{\rm s}^{-1}$. The error is calculated as the standard deviation of all values of $C(\sigma_{200})$ from each SSP as shown in Fig.~\ref{fig:correctionfactor}, when correcting from the maximal (central) velocity dispersion of each galaxy.}
  \label{tab:corrsigerrors}
  \begin{tabular}{@{}lccccc@{}}
  \hline
Galaxy & $\sigma_{\rm NaI_{SDSS}}$ & $\sigma_{\rm CaT}$ & $\sigma_{\rm MgI}$ & $\sigma_{\rm TiO}$ & $\sigma_{\rm FeH}$ \\
 \hline
 NGC4889 & 0.016 ($^{0.826}_{0.779}$) & 0.005 ($^{0.818}_{0.805}$) & 0.035 ($^{0.432}_{0.322}$) & 0.003 ($^{0.988}_{0.980}$) & 0.017 ($^{0.819}_{0.776}$)
 \vspace{3pt}\\
 NGC4874 & 0.007 ($^{0.943}_{0.922}$) & 0.002 ($^{0.960}_{0.955}$) & 0.014 ($^{0.763}_{0.719}$) & 0.001 ($^{0.996}_{0.994}$) & 0.007 ($^{0.950}_{0.930}$)
 \vspace{3pt}\\
 NGC4839 & 0.007 ($^{0.943}_{0.922}$) & 0.002 ($^{0.960}_{0.955}$) & 0.014 ($^{0.763}_{0.719}$) & 0.001 ($^{0.996}_{0.994}$) & 0.007 ($^{0.950}_{0.930}$)
 \vspace{3pt}\\
 NGC4873 &  0.002 ($^{1.020}_{1.013}$) & 0.000(4) ($^{1.006}_{1.005}$) & 0.004 ($^{1.076}_{1.065}$) & 0.000(2) ($^{1.001(1)}_{1.000(6)}$) & 0.002 ($^{1.017}_{1.011}$) \\
\hline
\end{tabular}
\end{table*}

\section{Telluric correction, sky subtraction, and Wing-Ford measurements}
\label{sec:fehsys}

\begin{figure*}
 \centering
 \includegraphics[width=8.5cm]{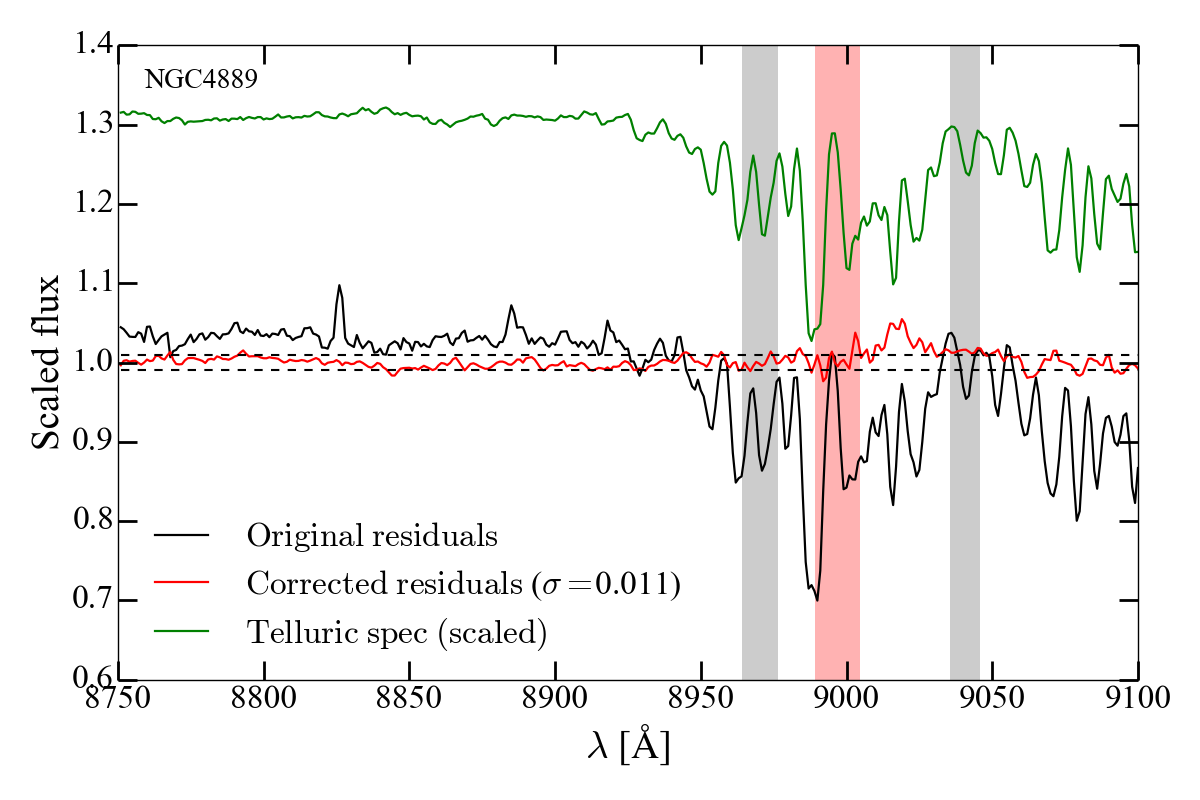}
  \includegraphics[width=8.5cm]{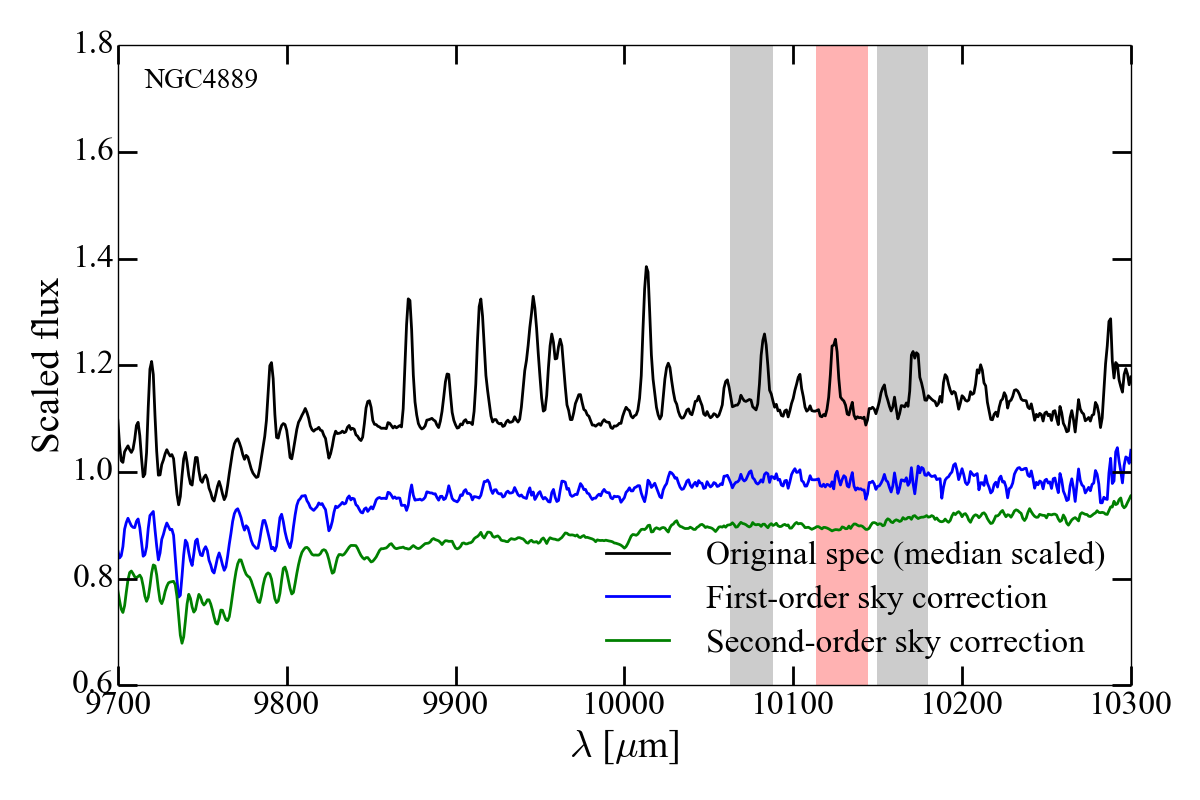} \\
   \includegraphics[width=8.5cm]{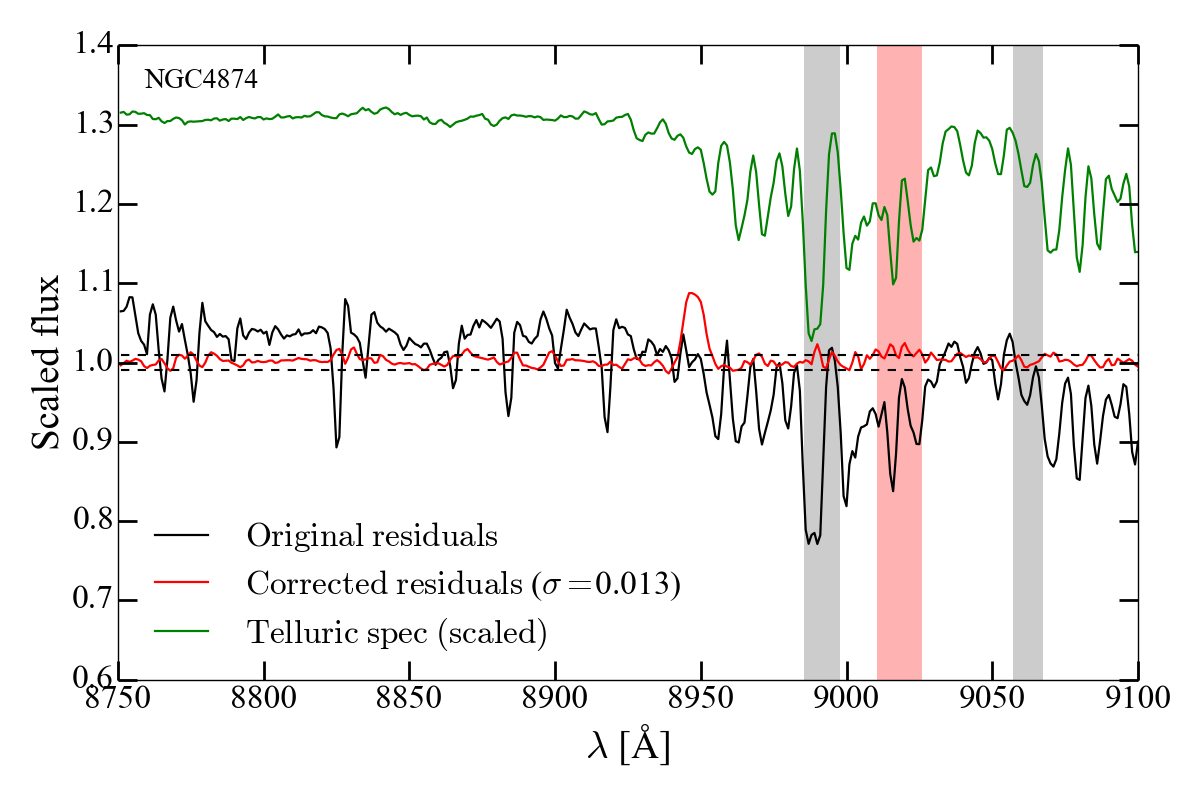}
  \includegraphics[width=8.5cm]{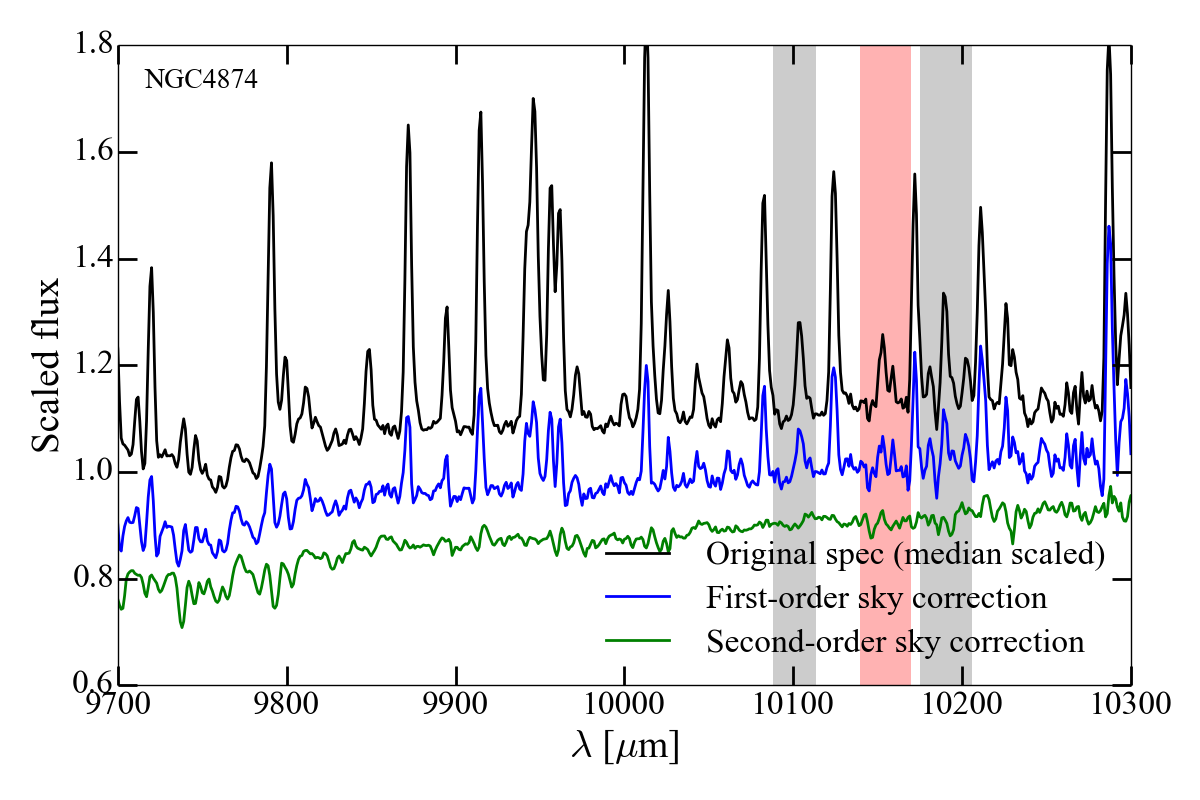} \\
   \includegraphics[width=8.5cm]{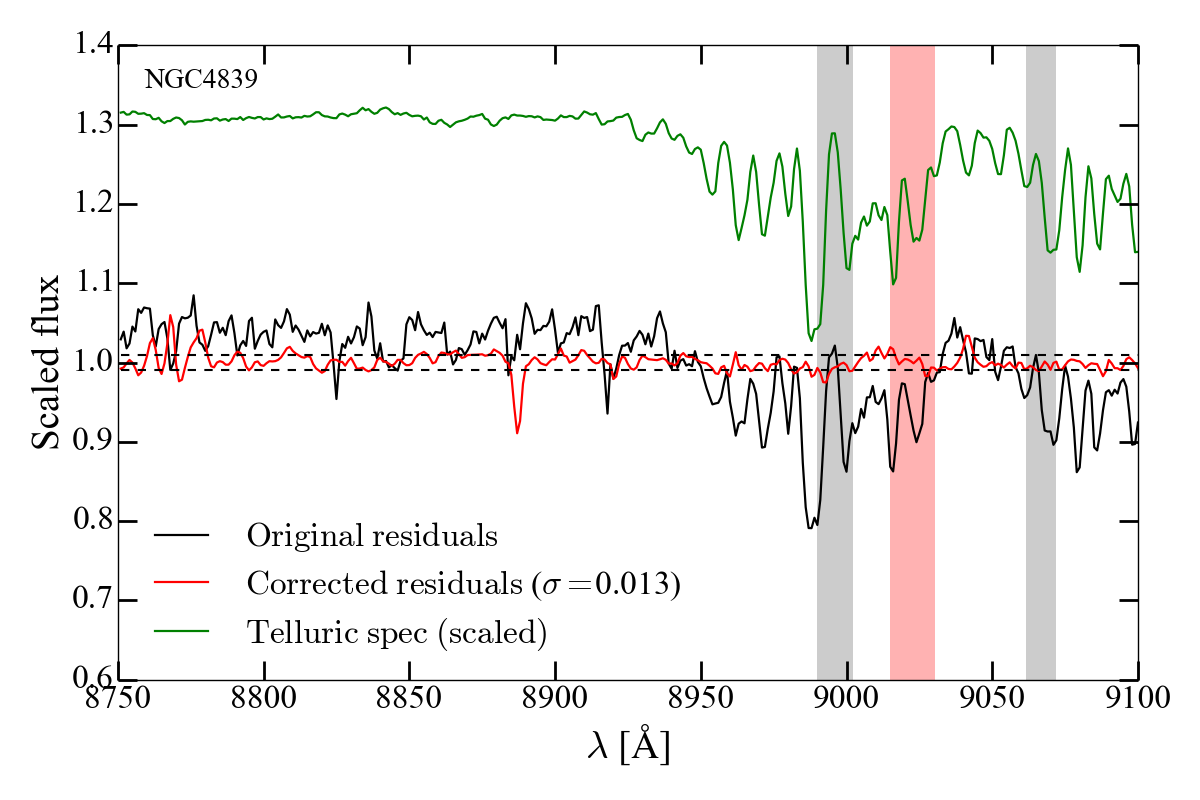}
  \includegraphics[width=8.5cm]{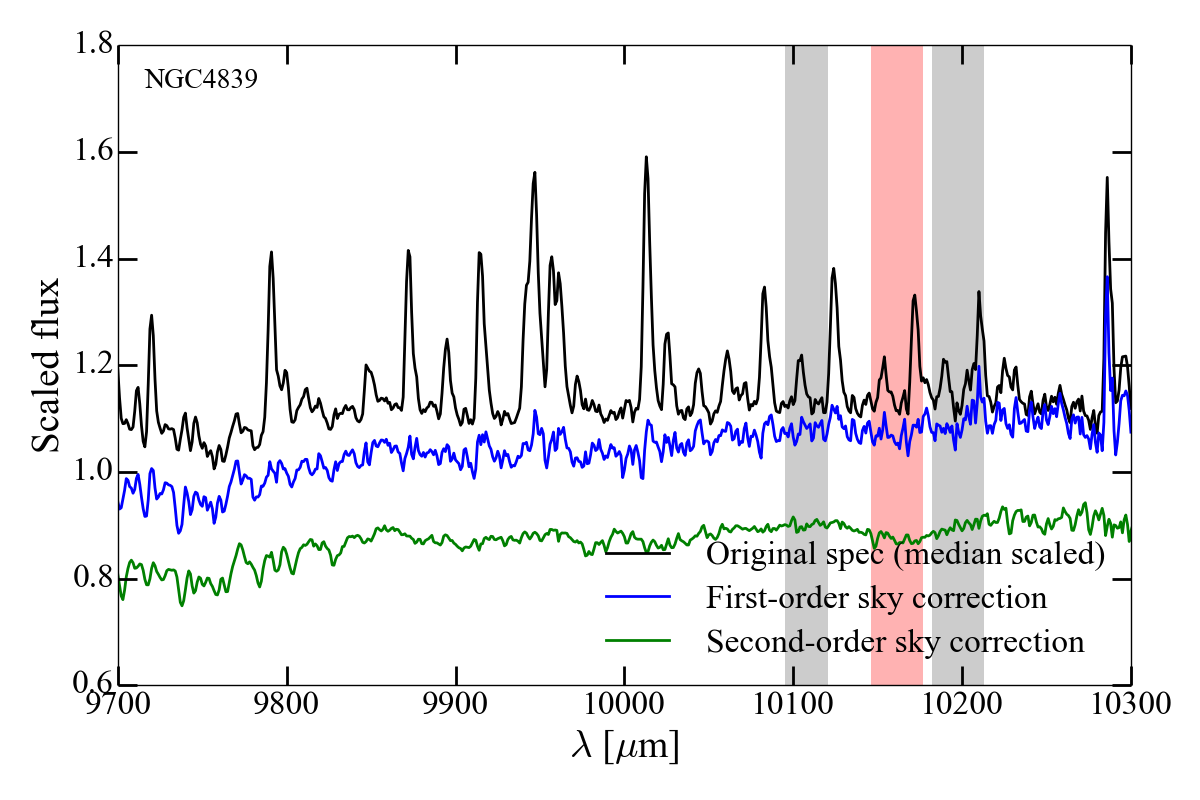} \\
 \caption{Plots showing telluric correction (left) and first- and second-order sky subtraction (right) for each BCG. The top two panels show NGC4889, the middle two show NGC4874 and the bottom two show NGC4839. The left-hand subplots show the telluric correction around MgI. The black line shows the ratio between original spectrum and kinematic template (assuming this is the intrinsic `telluric free' spectrum), the red line shows the same ratio with telluric division, and the green line shows the telluric spectrum (arbitrarily scaled to fit onto plot). The position of the MgI feature and pseudo-continuum is shown by the shaded regions. The right-hand subplots show a comparison of the first- and second-order sky subtractions around the FeH feature. The first-order subtraction is achieved using a static sky data cube of the same size as the science cube (with separate cubes correctly shifted as to match spaxel positions). The second-order correction is performed using {\sc ppxf} and removes the residuals at the level of a few percent left over from the first-order subtraction. The spectra are scaled by the median values and shifted vertically in the $y$-axis for clear presentation.}
 \label{fig:allgalstelluricsky}
\end{figure*}

We show here additional plots concerning the accuracy of our telluric correction and sky subtraction techniques. Fig.~\ref{fig:allgalstelluricsky} shows the telluric correction around MgI (left-hand side) and the {\sc ppxf} sky subtraction around FeH (right-hand side) for each of the BCGs; NGC4889 (top row), NGC4874 (middle row) and NGC4839 (bottom row). For each galaxy, the MgI region is the worst affected by telluric absorption and we show the residuals for this region. We find smaller residuals around the NaI$_{\rm{SDSS}}$, CaT and TiO features of no worse than 0.01. We do not incur residuals around the FeH feature as we divide by a smooth fit to the throughput curve instead of by a corrected telluric spectrum. The right hand plots show the first- and second-order sky subtractions on central spectra around FeH using {\sc ppxf}. Black lines show the original spectrum with no sky subtraction. Blue lines show the spectrum after first-order O-S cube sky subtraction; the first-order subtraction is achieved using a static sky data cube of the same size as the science cube (with separate cubes correctly shifted as to match spaxel positions). Green lines show the final spectrum after second-order {\sc ppxf} sky subtraction; the second-order correction is performed using {\sc ppxf} and removes the residuals at the level of a few percent left over from the first-order subtraction.

\begin{figure*}
 \centering
 \includegraphics[width=8.5cm]{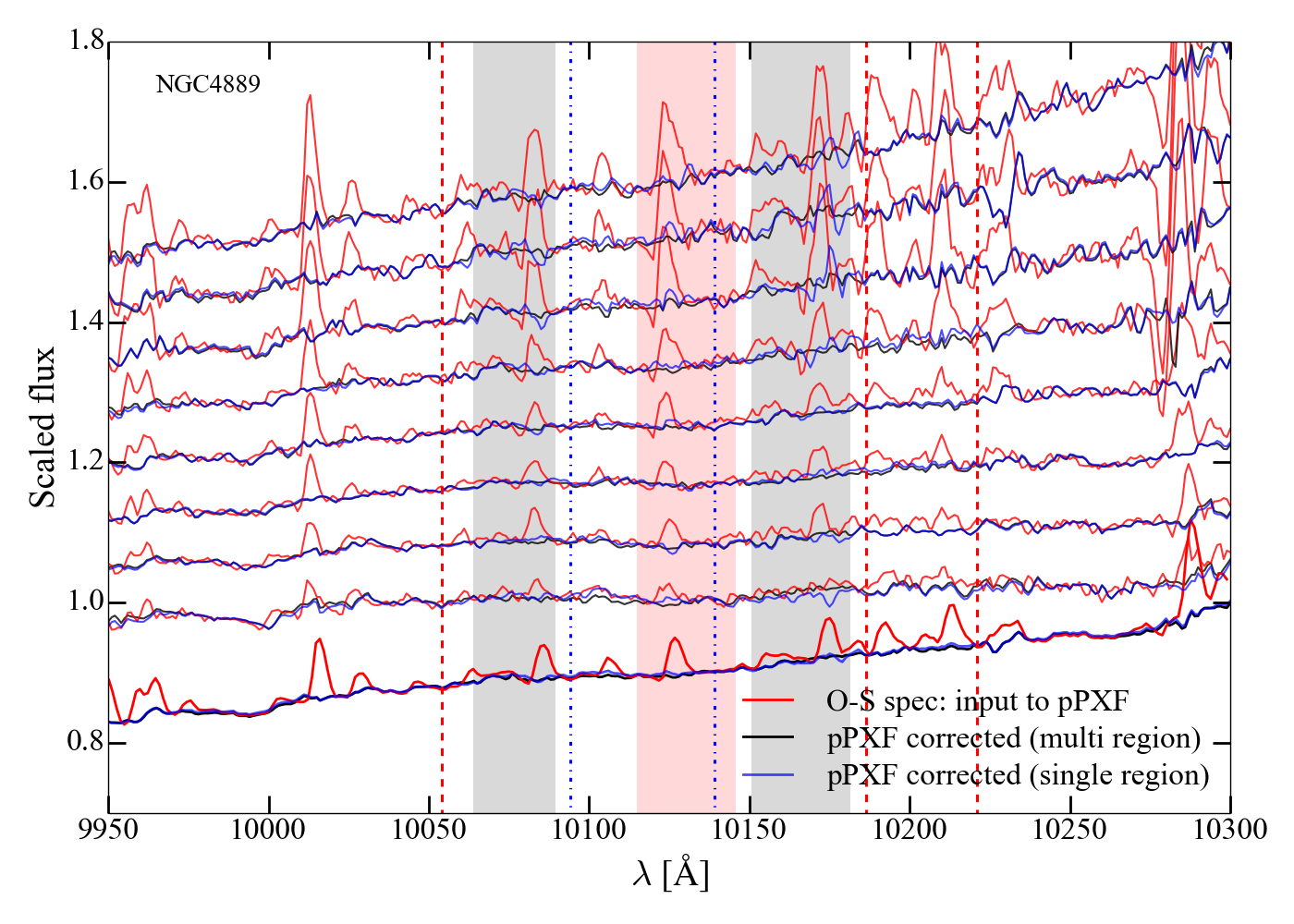}
  \includegraphics[width=8.5cm]{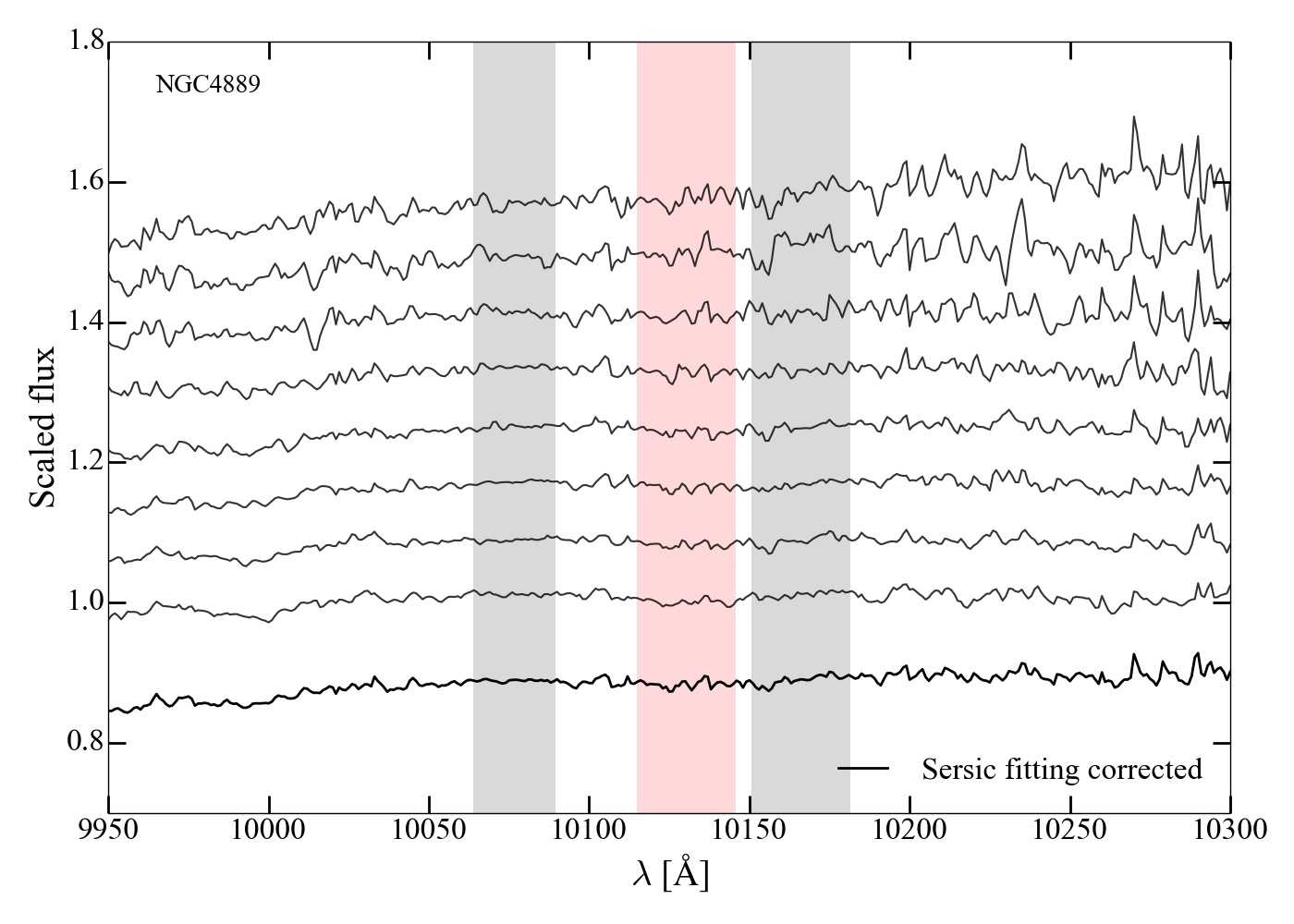} \\
   \includegraphics[width=8.5cm]{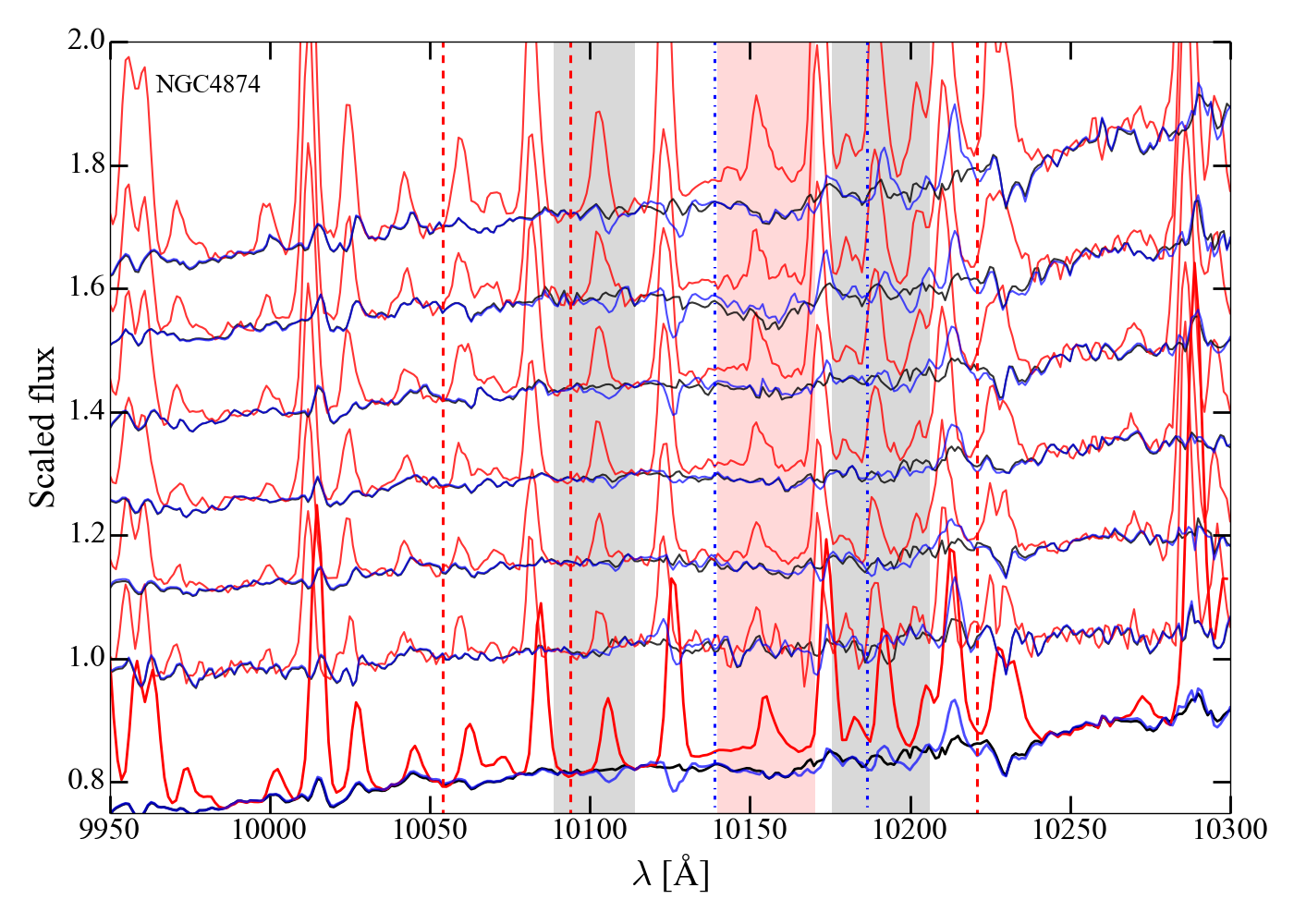}
  \includegraphics[width=8.5cm]{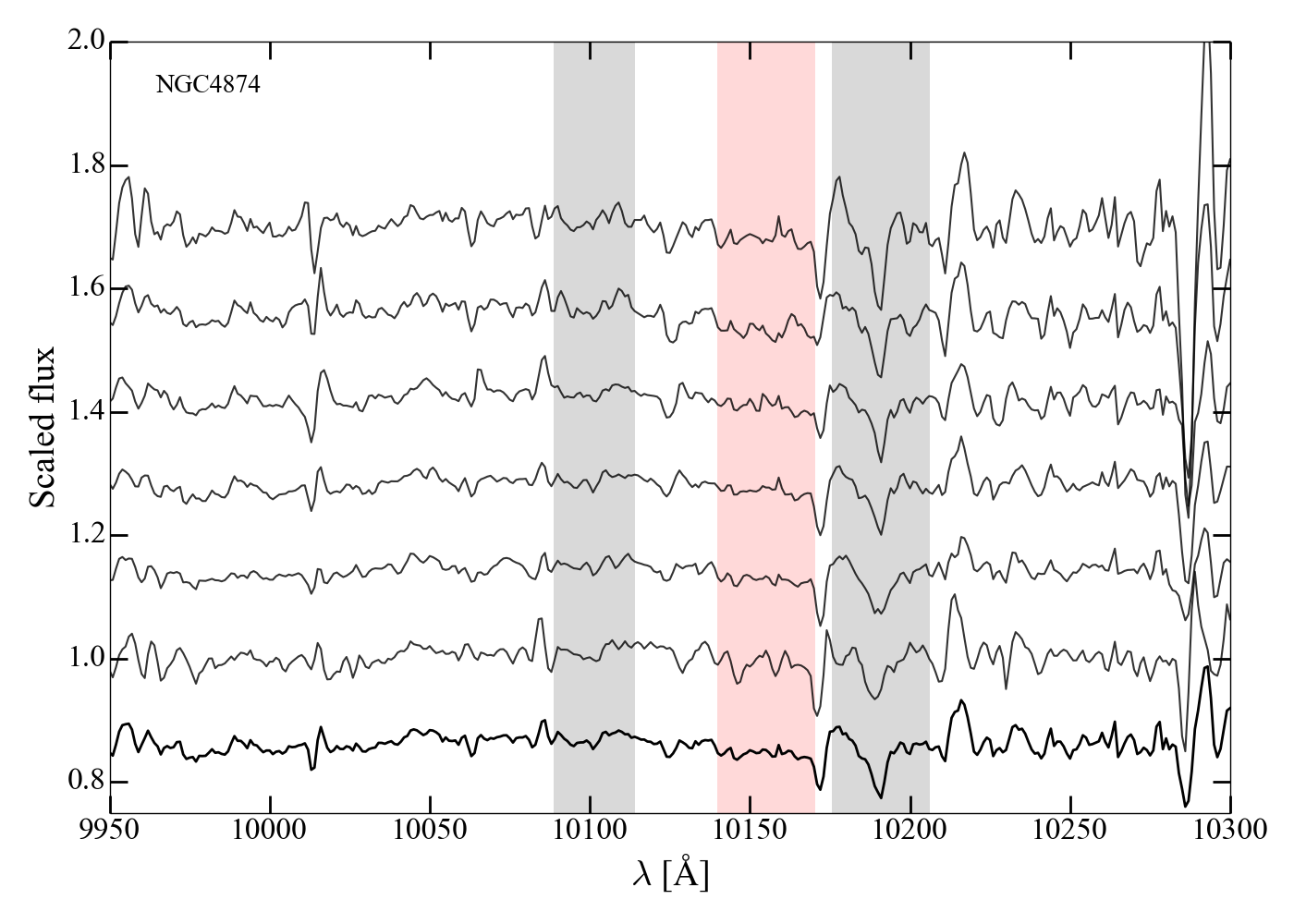} \\
   \includegraphics[width=8.5cm]{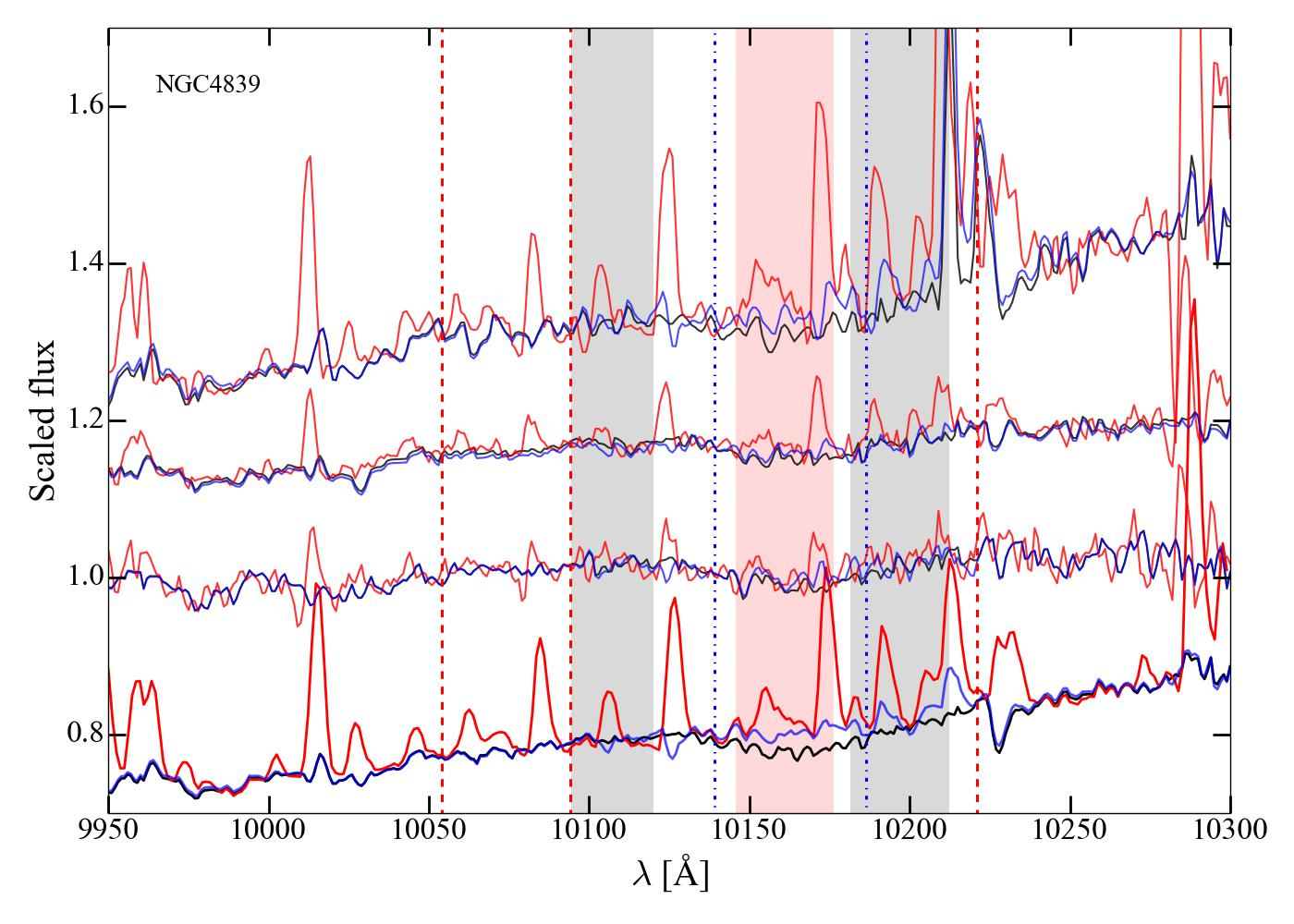}
  \includegraphics[width=8.5cm]{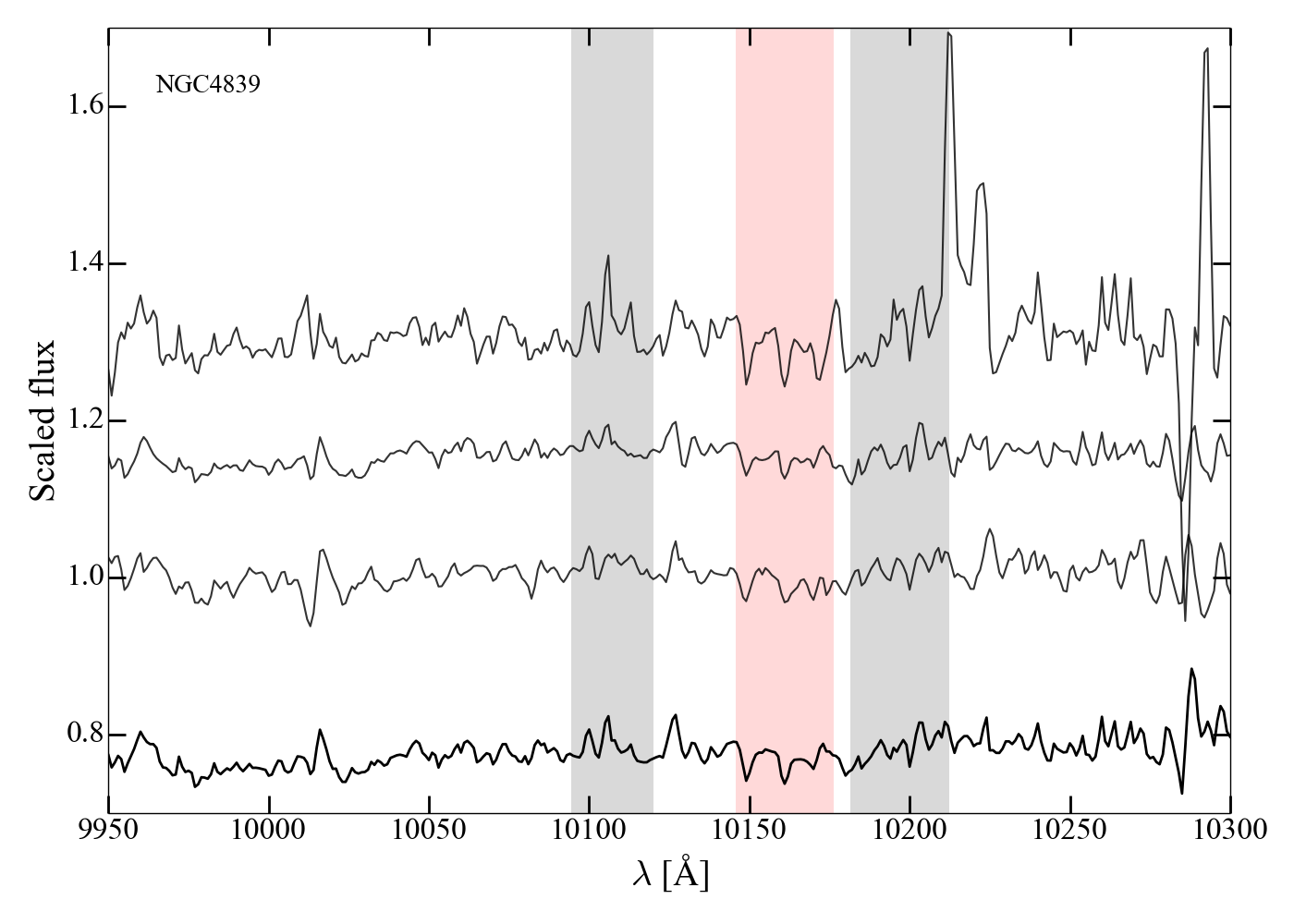} \\
 \caption{Plots showing the global (bold lines at bottom) and radial spectra (increasing radial distance bottom to top) around FeH for each BCG, before and after the two second-order sky subtraction routines: the {\sc ppxf} method on the left hand side, and the Sérsic profile fitting on the right hand side. The top two panels show NGC4889, the middle two show NGC4874 and the bottom two show NGC4839. On the left hand panels, the input spectra to {\sc ppxf} (red lines) are compared to the corrected spectra in black and blue. The sky regions that are scaled by {\sc ppxf} to fit the sky lines in the science spectra are shown by the vertical lines. The black line output spectra are the results when scaling all sky regions shown by the red and blue vertical lines. The blue output spectra are the results when removing the sky regions shown by the vertical blue dotted lines and {\it only scaling a single sky region over the FeH feature and pseudo-continuum definitions} (shaded regions: shown at the redshift for each BCG). The right hand plots show the spectra corrected by the Sérsic profile fitting method as a comparison. NGC4889 shows consistent spectra at all radii. NGC4874 and NGC4839 have much more prominent sky lines across the FeH feature due to the increased redshift. The global spectra for NGC4839 shows a large systematic difference between the multi- and single- sky region {\sc ppxf} correction, with the multi-region spectrum showing a much deeper feature.}
 \label{fig:allgalsfehspecs}
\end{figure*}

Fig.~\ref{fig:allgalsfehspecs} shows the global and radial spectra around FeH for each BCG, before and after the two second-order sky subtraction routines: the {\sc ppxf} method on the left hand side, and the Sérsic profile fitting on the right hand side. The top two panels show NGC4889, the middle two show NGC4874 and the bottom two show NGC4839. On the left hand panels, the input spectra to {\sc ppxf} (red lines) are compared to the corrected spectra in black and blue. The sky regions that are scaled by {\sc ppxf} to fit the sky lines in the science spectra are shown by the vertical lines. The black line output spectra are the results when scaling all sky regions shown by the red and blue vertical lines. The blue output spectra are the results when removing the sky regions shown by the vertical blue dashed-dotted lines and {\it only scaling a single sky region over the FeH feature and pseudo-continuum definitions} (shaded regions: shown at the redshift for each BCG). Only scaling a single sky region across the FeH feature stops any differential gradient being introduced that can alter the resulting feature measurement on the order of tens of per cent. The right hand plots show the spectra corrected by the Sérsic profile fitting method as a comparison. NGC4889 shows consistent spectra at all radii. NGC4874 and NGC4839 have much more prominent sky lines across the FeH feature due to the increased redshift. The global spectra for NGC4839 shows a large systematic difference between the multi- and single- sky region {\sc ppxf} correction, with the multi-region spectrum showing a much deeper feature. Thus, we cannot be certain whether the deeper FeH feature measured for NGC4839 is real or a result of the reduction process.

\begin{figure*}
 \centering
 \includegraphics[width=14cm]{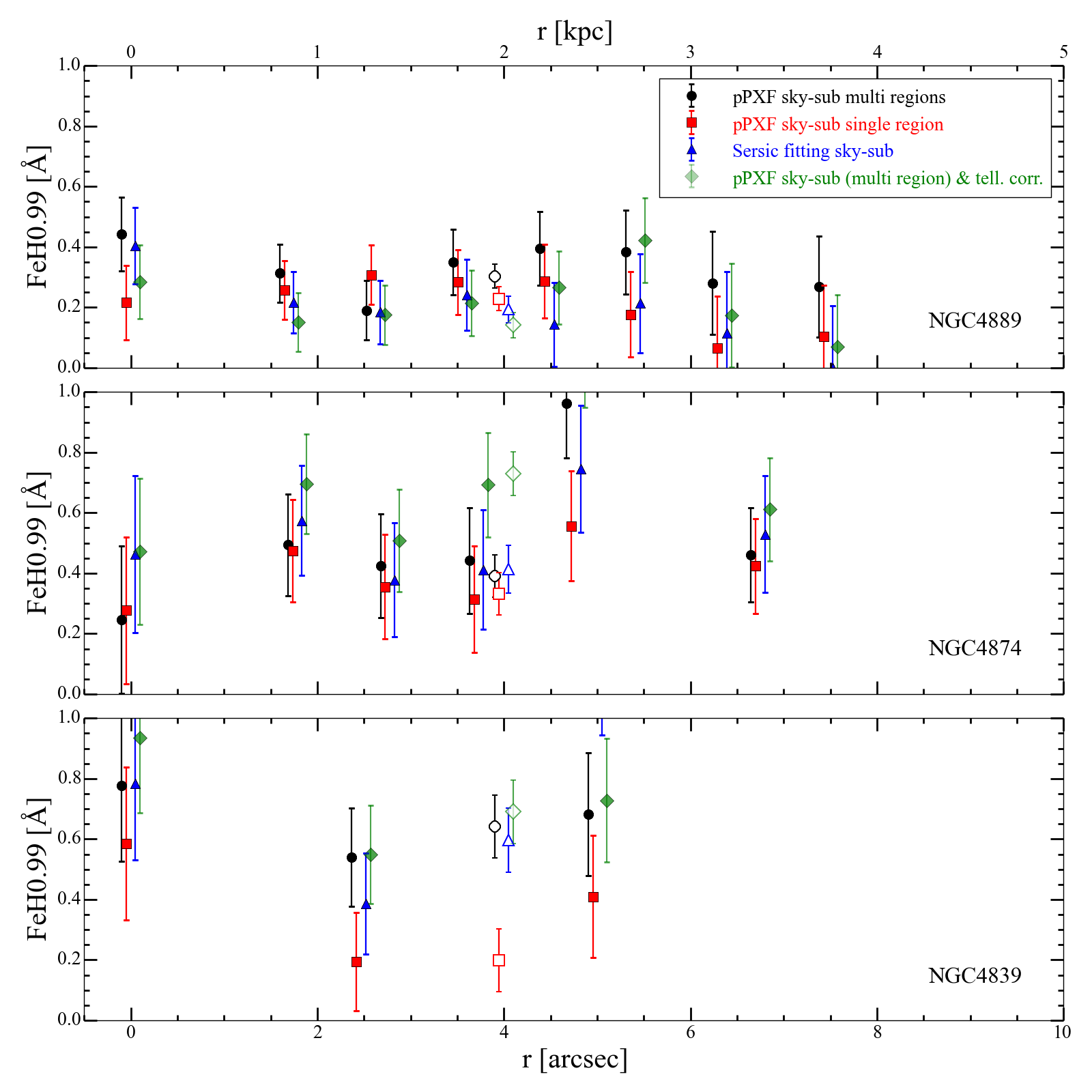}
 \caption{Plot showing radial measurements of FeH for each BCG, at a common resolution of $\sigma=200\,{\rm km}\,{\rm s}^{-1}$, after the different sky-subtraction and telluric correction procedures. The sub-panels are labelled for each galaxy with NGC4889 at the top, NGC4874 in the middle and NGC4839 at the bottom. In each subplot, the black circles show measurements from spectra corrected by telluric polynomial fit division, and sky subtracted using {\sc ppxf} with multiple sky regions across the FeH feature (shown in Fig.~\ref{fig:allgalsfehspecs}); the red squares are the same as the black circles but with only a single sky region across the FeH feature in the {\sc ppxf} sky subtraction; the blue triangles show the measurements from spectra corrected by telluric polynomial fit division, and sky subtracted using the Sérsic fitting procedure in Section~\ref{sec:sersic}; the green diamonds show measurements from spectra corrected using a telluric spectrum and sky subtracted using {\sc ppxf} with multiple sky regions across the FeH feature.}
 \label{fig:allFeH}
\end{figure*}

Fig.~\ref{fig:allFeH} shows our measurements of the FeH index for each BCG using the spectra from our three separate data reduction processes discussed in Section~\ref{sec:sersic}:\\
(a) telluric corrected spectra sky-subtracted using {\sc ppxf} (multiple sky regions: green diamonds),\\
(b) throughput corrected spectra sky-subtracted using {\sc ppxf} (multiple sky regions: black circles, single sky region: red squares),\\
(c) throughput corrected spectra sky-subtracted using O-S spatial fitting (blue triangles).\\
For NGC4889 all measurements are in good agreement showing weak FeH at all radii. For NGC4874 the telluric corrected spectrum results in a much deeper FeH strength compared with the telluric continuum fitting method. This is caused by systematics in removing the Paschen features from the telluric spectrum artificially deepening the FeH feature. We therefore caution against performing telluric correction when FeH is redshifted long-ward of around $1\,\rm{\mu m}$, as it moves into a region of negligible telluric absorption so any correction can introduce excess residuals. NGC4839 shows a large range of FeH strengths from the different sky subtraction procedures. The multiple sky region {\sc ppxf} subtraction produces systematically deeper FeH features compared to the single sky region spectra. From Fig.~\ref{fig:allgalsfehspecs} we see that FeH is redshifted into a prominent blend of sky lines between 1.015--1.025\AA, which is challenging to remove, and is therefore very difficult to measure cleanly. This causes the large systematic uncertainty on our FeH measurements for NGC4839, and also presents a limitation of the {\sc ppxf} subtraction routine, justifying our use of an independent subtraction method.

\section{Deriving IMF slopes and caveats}
\label{sec:imfcaveats}

To derive the IMF slopes we use the age-IMF grid of SSPs from the CvD12 models, covering 3-13.5 Gyr and bottom-light to bottom-heavy $x=3$ IMFs. Using the age of each galaxy inferred from Fig.~\ref{fig:lickindices} we interpolate the SSP grid and find a set of SSPs for the range of IMFs at a given age. We then use the varying element and $\alpha$-enhancement SSPs to derive response factors to apply to the varying IMF SSPs for variations in $[\alpha$/Fe], [Fe/H] and [Na/Fe]. By measuring the FeH index for each SSP (corrected for metallicity, $\alpha$- and Na-enhancement), we compare the measured {\it global} value of FeH for each galaxy to the range of FeH values from the varying IMF SSPs. We propagate the fractional uncertainty in the FeH measurement through to the uncertainty in IMF slope as a first-order error approximation. We approximate a Chabrier IMF using a unimodal slope of $x=1.8$ and a bottom-light IMF using $x=1.6$ based upon matching the $(M/L)_R$ with that of the V15 unimodal, solar metallicity, 12.6 Gyr SSPs.

\begin{figure*}
\centering
\includegraphics[width=16cm]{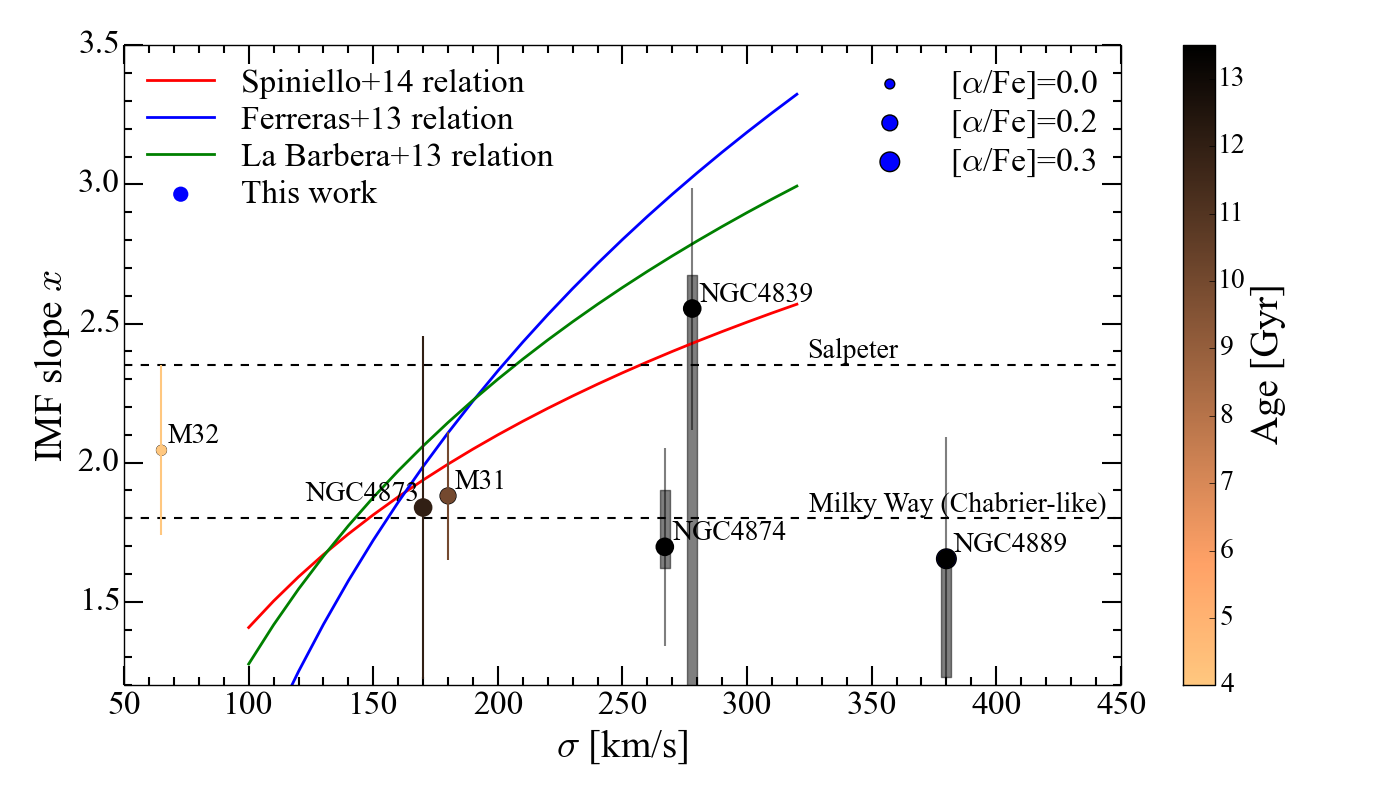}
\caption{Same as Fig.~\ref{all_galaxies_imf_constraints} but {\it without} response functions applied when deriving the IMF slope: Derived IMF slope $x$ against central velocity dispersion for our sample of Coma galaxies, as well as M32 and M31 from \citet{Zieleniewski2015}. The IMF slope has been derived from our {\it global} FeH measurements. For our data, the colour of each point indicates the age and the marker size indicates the level of [$\alpha$/Fe] enhancement for each galaxy. The coloured lines show {\bf unimodal} IMF-$\sigma_*$ relations from the previous works of \citet{Ferreras2013, LaBarbera2013} and \citet{Spiniello2014}, derived using central $2.7''$ SDSS spectra and several optical/NIR features, including NaD/NaI but excluding FeH. The dashed horizontal lines show the Salpeter and Milky-Way (Chabrier-like) slopes. Our results show consistency with a universal Chabrier IMF across a large range of velocity dispersions, with the exception of the single galaxy NGC4839, for which we see possible evidence of an IMF slightly heavier than Salpeter, albeit with a very large systematic uncertainty.}
\label{all_galaxies_imf_constraints_no-corr}
\end{figure*}

When deriving our IMF slopes as shown in Fig.~\ref{all_galaxies_imf_constraints} we apply response functions to the SSPs to account for variations in [$\alpha$/Fe], [Fe/H] and [Na/Fe]. The CvD12 models do not fully explore variations in total metallicity [Z/H] so the corrections we apply to the SSP spectra (before measuring FeH to derive the IMFs as shown in Fig.~\ref{all_galaxies_imf_constraints}), to account for variations in [$\alpha$/Fe], [Fe/H] and [Na/Fe], are approximations. In Fig.~\ref{all_galaxies_imf_constraints_no-corr} we show simple derivations of the IMF slope but {\it without} applying response functions to the SSPs to account for the stellar populations parameters. The results are fully consistent between the two plots showing that the variations in SSP parameters only yield second order effects. For example, increased metallicity, which enhances the FeH index strength, is offset by enhanced [Na/Fe], which decreases the FeH strength in the CvD12 models. A comprehensive set of SPS models that allows for changes in the key parameters of [Z/H], [$\alpha$/Fe], individual element abundances (especially sodium), and covers the FeH wavelength region will help provide more accurate constraints on the IMF in ETGs.

\bsp

\label{lastpage}

\end{document}